\documentclass[sigconf, nonacm,hyphens,7pt]{acmart}
\usepackage{tikz}
\newcommand*\circled[2][]{\tikz[baseline=(char.base)]{
    \node[shape=circle,draw,inner sep=1pt,#1,every number/.try] (char) {\color{black}#2};}}
\tikzstyle{every number}=[draw=black,fill=white] 
\usepackage{xcolor}
\usepackage{xspace}
\usepackage{color}
\usepackage{paralist}
\usepackage[font=small]{caption}
\definecolor{vegablue}{HTML}{4C78A8}
\definecolor{vegaorange}{HTML}{f58517}
\definecolor{vegared}{HTML}{e45755}
\definecolor{vegacyan}{HTML}{72b7b2}
\definecolor{vegagreen}{HTML}{54a24b}

\usepackage{cleveref}

\crefformat{section}{\S#2#1#3} 
\crefformat{subsection}{\S#2#1#3}
\crefformat{subsubsection}{\S#2#1#3}

\DeclareRobustCommand{\luxfig}{%
  \begingroup\normalfont
  \raisebox{-0.5ex}{\includegraphics[height=1.8\fontcharht\font`\B]{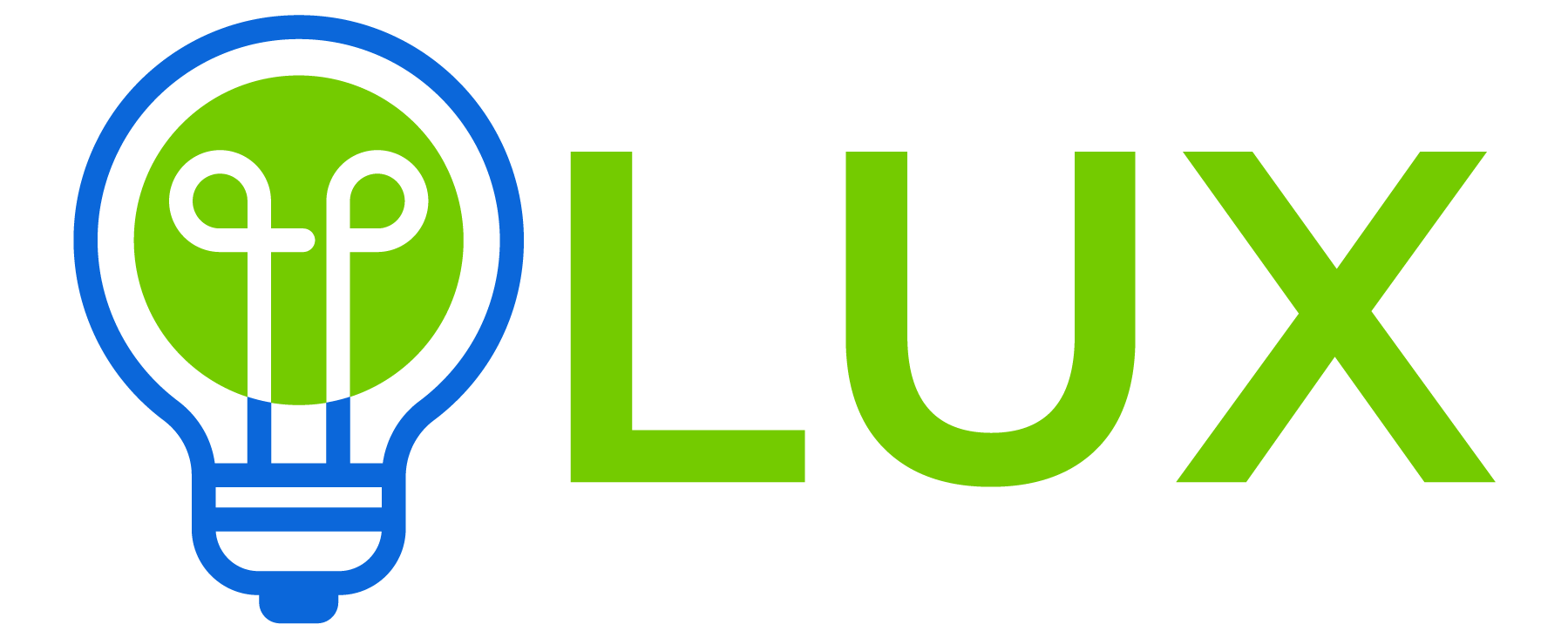}}%
  \endgroup
}

\newcommand{\verbat}[1]{{\small \begin{verbatim} #1 \end{verbatim}}}

\newcommand{\stitle}[1]{\smallskip

\noindent\textbf{#1}}
\newcommand{\emtitle}[1]{\vspace{0.8pt}

\noindent{\it #1}}
\newcommand{\utitle}[1]{\smallskip

\noindent{\it #1}}

\newcommand{\query}[1]{\par\noindent\textsc{#1.}}
\newcommand{\npar}{\par\noindent}
\newcommand{\cut}[1]{}
\newcommand{\tr}[1]{#1}
\newcommand{\papertext}[1]{}
\newcommand{\ccut}[1]{} 

\newcommand{\doris}[1]{\textcolor{violet}{[Doris: #1]}} 
\newcommand{\dtc}[1]{\textcolor{red}{[DT:#1]}}

\newcommand{\change}[1]{#1}

\newcommand{\code}[1]{\texttt{#1}\xspace}
\newcommand{\lux}{\textsc{Lux}\xspace}
\newcommand{\pandas}{\texttt{pandas}\xspace}
\newcommand{\airbnb}{\texttt{Airbnb}\xspace}
\newcommand{\communities}{\texttt{Communities}\xspace}
\newcommand{\mpl}{\texttt{matplotlib}\xspace}
\newcommand{\clause}{\texttt{Clause}\xspace}
\newcommand{\clauses}{\texttt{Clause}s\xspace}
\newcommand{\axis}{\texttt{Axis}\xspace}
\newcommand{\filter}{\texttt{Filter}\xspace}
\newcommand{\vis}{\texttt{Vis}\xspace}
\newcommand{\vislist}{\texttt{VisList}\xspace}
\newcommand{\luxdf}{\texttt{LuxDataFrame}\xspace}

\newcommand{\visrec}{VisRec\xspace}
\newcommand{\visspec}{VisSpec\xspace}

\newcommand{\wflow}{\textsc{wflow}\xspace}
\newcommand{\prune}{\textsc{prune}\xspace}
\newcommand{\async}{\textsc{async}\xspace}
\newenvironment{denselist}{
    \begin{list}{\tiny{$\bullet$}}%
    {\setlength{\itemsep}{0ex} \setlength{\topsep}{0ex}
    \setlength{\parsep}{0pt} \setlength{\itemindent}{0pt}
    \setlength{\leftmargin}{1em}
    \setlength{\partopsep}{0pt}}}%
    {\end{list}}

\newcommand{\removed}[1]{} 

\newcommand\vldbdoi{10.14778/3494124.3494151}
\newcommand\vldbpages{727 - 738}
\newcommand\vldbvolume{15}
\newcommand\vldbissue{3}
\newcommand\vldbyear{2022}
\newcommand\vldbauthors{Doris Jung-Lin Lee, Dixin Tang, Kunal Agarwal, Thyne Boonmark, Caitlyn Chen, Jake Kang, Ujjaini Mukhopadhyay, Jerry Song, Micah Yong, Marti A. Hearst, Aditya G. Parameswaran}
\newcommand\vldbtitle{Lux: Always-on Visualization Recommendations for Exploratory Dataframe Workflows} 
\newcommand\vldbavailabilityurl{\url{https://github.com/lux-org/lux}}
\newcommand\vldbpagestyle{empty}
 
\begin{document}
\title{Lux: Always-on Visualization Recommendations \\for Exploratory Dataframe Workflows \\\tr{{\LARGE[Technical Report]}}}
\author{Doris Jung-Lin Lee, Dixin Tang, Kunal Agarwal, Thyne Boonmark, Caitlyn Chen, Jake Kang, Ujjaini Mukhopadhyay, Jerry Song, Micah Yong, Marti A. Hearst, Aditya G. Parameswaran}
\affiliation{\textbf{UC Berkeley}}
\email{{dorislee,totemtang,kagarwal2,thyneboonmark,caitlynachen,cjache,ujjaini,jerrysong,micahtyong,hearst,adityagp}@berkeley.edu}

\begingroup
 \def\UrlFont{\small} 
 \mathchardef\UrlBreakPenalty=10000

\begin{abstract}  
  Exploratory data science largely happens in 
  computational notebooks with dataframe API\change{s}, such as \pandas, that support flexible means to transform, clean, and analyze data.
  Yet, visually exploring data in dataframes remains tedious, requiring substantial programming effort for visualization 
  and mental effort to determine what analysis to perform next. We propose \lux, an \emph{always-on} framework for accelerating visual insight discovery in \change{dataframe} workflows. When users print a dataframe in their notebooks, \lux recommends visualizations to provide a quick overview of the patterns and trends and suggests promising analysis directions. \lux features a high-level language for generating visualizations on demand to encourage rapid visual experimentation with data. We demonstrate that through the use of a careful design and three system optimizations, \lux 
  adds no more than two seconds of overhead
  on top of \pandas for over 98\% of datasets in the UCI repository. We evaluate \lux in terms of usability via\tr{ a controlled first-use study and} interviews with early adopters, finding that \lux helps fulfill the needs of data scientists for visualization support within their dataframe workflows. \lux has already been embraced by data science practitioners, with over \change{3.1k stars on Github.}
  \end{abstract}
\maketitle

\pagestyle{\vldbpagestyle}
\begingroup\small\noindent\raggedright\textbf{PVLDB Reference Format:}\\
\vldbauthors. \vldbtitle. PVLDB, \vldbvolume(\vldbissue): \vldbpages, \vldbyear.\\
\href{https://doi.org/\vldbdoi}{doi:\vldbdoi}
\endgroup
\begingroup
\renewcommand\thefootnote{}\footnote{\noindent
This work is licensed under the Creative Commons BY-NC-ND 4.0 International License. Visit https://creativecommons.org/licenses/by-nc-nd/4.0/ to view a copy of this license. For any use beyond those covered by this license, obtain permission by emailing \href{mailto:info@vldb.org}{info@vldb.org}. Copyright is held by the owner/author(s). Publication rights licensed to the VLDB Endowment. \\
\raggedright Proceedings of the VLDB Endowment, Vol. \vldbvolume, No. \vldbissue\ %
ISSN 2150-8097. \\
\href{https://doi.org/\vldbdoi}{doi:\vldbdoi} \\
}\addtocounter{footnote}{-1}\endgroup

\ifdefempty{\vldbavailabilityurl}{}{
\vspace{.3cm}
\begingroup\small\noindent\raggedright\textbf{PVLDB Artifact Availability:}\\
The source code, data, and/or other artifacts have been made available at \url{https://github.com/lux-org/lux}.
\endgroup
}

\thispagestyle{plain}
\pagestyle{plain}
\pagenumbering{arabic}

\section{Introduction}~\label{sec:intro}

\change{Exploratory data science} is an iterative, trial-and-error process,
involving many interleaved stages of data cleaning,
transformation, analysis, and visualization. 
Data scientists typically use a dataframe library~\cite{modin-vision,jindal2021magpie},
such as \pandas~\cite{reback2020pandas},
which offers a flexible and rich set of operators
to transform, analyze, and clean tabular datasets.
They manipulate dataframes within a computational
notebook such as Jupyter, which offers a flexible medium
to write and execute snippets of code; 
nearly 75\% of data scientists 
use them everyday~\cite{kagglesurvey2020}. 
In between these dataframe transformation operations, 
users visually inspect intermediate results,
either by printing the dataframe, 
or by using a visualization library
to generate visual summaries.
This visual inspection is {\em essential} to validate
whether the prior operations had their desired
effect and determine what needs to be done next. 
However, {\bf \em visualizing 
dataframes is an unwieldy and error-prone process, 
adding substantial friction to the fluid, iterative
process of data science}, for two reasons:
cumbersome boilerplate code and
challenges in determining the next steps.

\stitle{Cumbersome Boilerplate Code.}
Substantial
boilerplate code is 
necessary to simply generate a visualization
from dataframes.
In a formative study,
we analyzed a sample of 587 publicly-available
notebooks from Rule et al.~\cite{Rule2018}
to understand current visualization practices.
A surprising number of notebooks 
apply a series of {\em data processing} operations 
to wrangle the dataframe into a form amenable 
to visualization, followed by a set of 
highly-templatized {\em visualization specification} 
code snippets copy-and-pasted across the notebook.
Our findings echo a recent study
of 6386 Github notebooks~\cite{copypastenotebook},
where visualization code was the most dominant category of
duplicated code (21\%).
On top of the high cognitive cost 
when writing ``glue code'' 
to go from dataframes to visualizations~\cite{alspaugh2019,Wang_Viser2020}, 
users have to context-switch between thinking 
about data operations and visual elements. 
These barriers hinder exploratory visualizations and, 
as a result, users often only visualize 
during the ``\emph{late stages of [their] workflow}''~\cite{Batch2018,Kandel2012Interview}, 
rather than for experimenting with 
possible analyses---which is precisely 
when visualization is likely to be most useful.

\stitle{Challenges in Determining Next Steps.}
Beyond writing code to generate a given visualization, 
there are challenges in determining which visualizations
to generate in the first place.
Dataframe APIs support datasets with 
millions of records and hundreds of attributes,
leading to many combinations of visualizations
that can be generated.
The many choices make
it hard for the
data scientist to determine
what visualization to generate to advance analysis, and automated assistance is not provided.

\stitle{Always-On \change{Dataframe Visualizations} with \luxfig.}
To address the above challenges, we introduce \lux,
a seamless extension to \pandas that retains its convenient and powerful API,
but enhances the tabular outputs with automatically-generated 
visualizations highlighting interesting
patterns and suggesting next steps for analysis.
\lux has already been adopted by data scientists 
from a diverse set of industries, and 
has gained traction in the open-source community, \change{with the number of {\bf \em monthly downloads around 9k
(with a total of 62k downloads), and over 3.1k stars on Github,
as of November 2021}.} 
Multiple industry users have created blog posts or YouTube videos extolling the virtues
of \lux~\cite{lux_article1,lux_video1,lux_video2,lux_article2,lux_article3,lux_article4}.
\change{
   \stitle{Contributions.} 
   Our contributions are as follows.

   First, we introduce a novel, always-on framework that provides
   visualizations for the dataframe 
   as it stands at any point in the workflow (\cref{sec:demo}). 
   This is in contrast with existing 
   visualization specification libraries~\cite{matplotlib,altair,ggplot} 
   that require users to write substantial code to generate visualizations. 
   This multi-tiered dataframe interaction framework supports \pandas' 600+ operators 
   without compromising the ease and flexibility of 
   data transformation and analysis (\cref{sec:framework}).
   
    Second, we introduce an expressive  and succinct {\em intent} language 
    powered by a formal,  algebra 
    that allows users to specify their fuzzy
    intent at a high level. 
    Compared to existing 
    languages\tr{ for partial  
    specification}~\cite{Lin2020,moritz2019formalizing,Siddiqui2017,wongsuphasawat2016towards}, 
    the intent language in \lux 
    not only allows users to create one or more 
    visualizations but 
    also flexibly indicate their high-level analysis interest, 
    without worrying about how the data elements map onto 
    aspects of the visualization (\cref{sec:intent}).

    Third, we introduce a novel recommendation system 
    that uses automatically extracted information 
    about dataframes to implicitly infer the appropriate visualizations to recommend. 
   This is in contrast with most existing visualization recommendation systems, 
   which are situated in GUI-based charting tools, whereas \lux is one of the first 
   of such systems that is designed to fit into a programmatic dataframe workflow. In particular, we introduce two novel classes of recommendations based on dataframe structure and history specific to such workflows (\cref{sec:rec}).
    
    Fourth, we identify opportunities wherein 
    we can adapt techniques from 
    approximate query processing~\cite{GarofalakisG01, CormodeGHJ12}, 
    early pruning~\cite{Vartak2015,kim2015rapid,fastmatch}, 
    caching and reuse~\cite{GuptaM05, TangSEKF19},
    and asynchronous computation~\cite{Xin2021,BendreWMCP19} 
    to provide interactive feedback, which is critical for usability; 
    \lux adds no more than two seconds of overhead 
    on top of medium-to-large 
    real-world datasets (\cref{sec:compute}).
    
    Finally, we evaluate the interactive latency of \lux 
    (\cref{sec:evaluation}) and usability
    with early adopters (\cref{sec:user}) 
    that assess the effectiveness of this lightweight, always-on approach to visualizing dataframes.
    }

\section{Related Work}\label{sec:relatedwork}

\lux draws from work on visualization recommendation systems, visualization specification, and visual dataframe tools.

\stitle{Visualization Recommendation (\visrec).}
To visualize data, data scientists
need to subselect the aspects 
of data, and then define
a mapping from data to graphical
encodings. Interactive interfaces, such as 
Tableau~\cite{tableau,Stolte2002} and 
PowerBI~\cite{powerbi},
offer easy-to-use 
interfaces 
for visualization construction. 
Some systems offer 
suggestions on other 
possible visualizations 
for users to browse through,
as visualization recommendations.
\visrec systems
can either suggest interesting portions
of the data to visualize based
on statistical properties~\cite{Vartak2015,Vartak2017,Lee2018,demiralp2017foresight,Siddiqui2017,Kamat2016TrendQuery,Mannino2018} 
or better ways to 
visualize attributes 
that users have selected~\cite{wongsuphasawat2016towards,Mackinlay1986,Mackinlay2007,moritz2019formalizing,vizml}. 
Similarly, there has been research on 
recommending interesting 
 attributes 
or filters to avoid 
manual data exploration during OLAP~\cite{Wu2013,Sarawagi2000, Lee:2019,Sarawagi1999,Sarawagi1998,Joglekar2015}. 
While interactive GUI-based tools have gained
adoption among business analysts, they
are not as widely used by data scientists
with programming expertise,
due to their lack of customizability
and integration with the rest of the
data science workflow. 
\lux draws on recommendation principles from this literature 
and explores how visualization recommendations can support a dataframe workflow.
Moreover, Figure~\ref{fig:framework} outlines a novel, multi-tiered framework that \lux employs to support flexible visual and programmatic interactions with a dataframe, overcoming the limitation in expressiveness of existing GUI-based \visrec tools.

\stitle{Visualization Specification (\visspec).}
\visspec frameworks codify 
visualization design principles 
and best practices to simplify 
the task of creating a visualization~\cite{satyanarayan2016reactive,satyanarayan2017vega,ggplot,Bostock2011,stolte2002polaris}. These frameworks encompass a range of abstractions
depending on the degree to which users are required 
to specify low-level details 
associated with the visualization definition. 
For example, \emph{imperative} visualization libraries, 
such as plotly~\cite{plotly}, D3~\cite{Bostock2011},
and matplotlib~\cite{matplotlib}, require users 
to manually compute the data associated 
with the graphical elements (e.g., position or size of marks) 
before defining the visualization characteristics. 
\emph{Declarative} visualization languages, 
such as Altair~\cite{altair} and Vega-Lite~\cite{satyanarayan2017vega}, 
enable rapid specification of visualizations 
by applying smart defaults to synthesize low-level visualization details, 
so that users are not required to specify common chart components, such as axes, ticks, 
and labels. \lux is built on top of these imperative 
and declarative frameworks and synthesizes visualization code 
to enable users to customize as needed.
\par \emph{Partial} specification languages, 
such as Draco~\cite{moritz2019formalizing} and 
CompassQL~\cite{wongsuphasawat2016towards}, 
commonly employed in \visrec systems, 
support reasoning based on a partial specification 
provided by the user and design constraints encoded 
in the system. 
A partial specification can be thought of as a ``query'', 
with the system automatically ranking a set of 
perceptually-effective visualizations that match the query. 
As we will see in Section~\ref{sec:intent}, the intent language in \lux is more convenient to specify than these existing languages in that it only requires users to specify data aspects of interest (or omit them entirely), instead of having to worry about visualization encodings. \lux is also more versatile in that it supports functionalities beyond visualization creation \change{for steering the recommendations generated}. That said, as a promising direction for future work, \lux could make use of Draco's sophisticated reasoning around visualization design to improve which visualizations are displayed, going beyond the rule-based heuristics in its current implementation.
\par Compared to imperative, declarative, and partial \visspec frameworks, 
Figure~\ref{fig:comparison} illustrates how \lux's intent language further reduces the specification burden on users, allowing them to provide lightweight intent as opposed to writing long 
code fragments for visualization; we will elaborate on this in Section~\ref{sec:intent}.
\stitle{Visual Data Exploration with Dataframes.}
Of late, dataframes have become
the de-facto framework for
interactive data science.
The comprehensive, incremental
set of operators make it easy
to do sophisticated data transformation, 
while also allowing validation
after each step.
However, 
exploring dataframes
is challenging,
requiring substantial programming
and analytical know-how.
Many visualization tools 
have been developed for dataframes~\cite{bamboolib,pandas-profiling,dataprep,sweetviz,pandasgui}. 
These tools generate 
summaries, 
covering analyses spanning 
missing values, outliers, 
attribute-level visualizations, 
and associated statistics. 
In addition, 
bamboo\-lib~\cite{bamboolib}\tr{, pandas-profiling~\cite{pandas-profiling}, dataprep~\cite{dataprep}, sweetviz~\cite{sweetviz},} 
and pandasgui~\cite{pandasgui} 
offer a GUI 
for constructing visualizations 
and data transformations. Unlike these existing tools, \lux lowers the barrier to visualizing dataframes by adopting an always-on approach so that dataframe visualizations are always recommended to users at all times\tr{, instead of relying on users to 
explicitly call external commands 
to \textit{plot} or \textit{profile} as needed}. 

\change{\section{Example Workfow}\label{sec:demo} 
In this example workflow, we} demonstrate 
how always-on visualization support
for dataframes accelerates
exploration and discovery. We present a workflow of Alice, a public policy analyst, exploring the relationship between world developmental indicators (such as life expectancy, inequality, and wellbeing) and the country's early effort in COVID-19 response. A live demo of the example notebook can be found at {\normalsize \url{http://tinyurl.com/demo-lux}}.

\stitle{Always-on dataframe visualization.} 
Alice opens up a Jupyter notebook 
and imports \pandas and \lux. 
Using \pandas's \texttt{read\_csv} command, 
Alice loads the Happy Planet Index (HPI)~\cite{hpi}
dataset of country-level data on sustainability and well-being. 
To get an overview, Alice 
prints\footnote{\scriptsize We refer to any operations 
that result in a dataframe in 
the output cell of a notebook 
as \textit{printing the dataframe}, 
not the literal `print (df)'.} 
the dataframe \texttt{df} and 
\lux displays 
the default \pandas tabular view, 
as shown in Figure~\ref{fig:print_df} (top, orange box). 
By clicking on the toggle button, 
Alice switches to the \lux view 
that displays a set of univariate 
and bivariate visualizations (bottom),
including scatterplots, bar charts, 
and maps, showing an overview of the trends. 
Visualizations are organized 
into sets called \textit{actions}, 
displayed as tabs.
The one displayed currently is the \code{Geographic}
action. 
By inspecting the \code{Correlation} 
tab in Figure~\ref{fig:print_df} (not displayed here), 
she learns that there is a negative correlation between \texttt{AvrgLifeExpectancy} and \texttt{Inequality} 
(same chart as Figure~\ref{fig:set_intent} left); in other words, countries with higher levels of inequality also have a lower average life expectancy.
She also examines the other tabs, 
which show the \code{Distribution} of quantitative attributes 
and the \code{Occurrence} of categorical attributes. 

\begin{figure}[h!]
    \includegraphics[width=\linewidth]{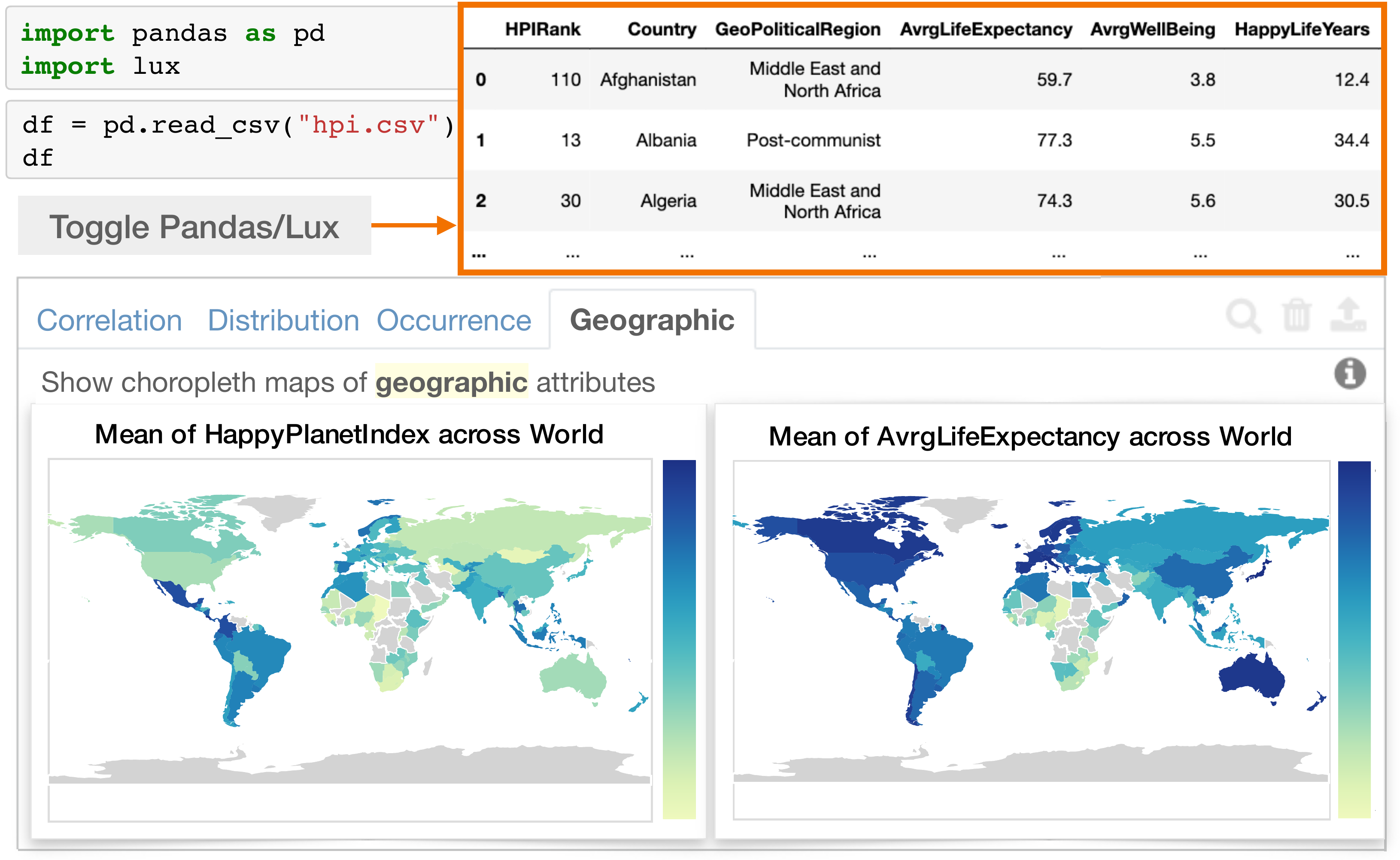}
    \caption{By printing out the dataframe, 
    the default \pandas tabular view 
    is displayed (orange box) 
    and users can toggle to browse 
    through visualizations recommended by \lux.}
    \label{fig:print_df}
\end{figure}
\stitle{Steering analysis with intent.} 
Next, Alice wants to investigate 
if any country-level characteristics explain the observed negative correlation between inequality and life expectancy.
As in Figure~\ref{fig:set_intent}, 
she specifies her analysis \textit{intent} 
to \lux as: \texttt{{\small df.intent = ["AvrgLifeExpectancy","Inequality"]}}. On printing the dataframe again, 
\lux employs the specified analysis intent 
to steer the recommendations towards what 
Alice might be interested in. 
On the left, Alice sees the visualization 
based on her specified intent. 
On the right, Alice sees two sets of 
recommendations that add an additional attribute (\code{Enhance}) 
or add an additional filter (\code{Filter}) 
to her intent. 
By looking at the colored scatterplots 
in the \code{Enhance} action, 
she learns that most G10 industrialized countries (Figure~\ref{fig:set_intent} center) 
are on the upper left quadrant on the scatterplot (low inequality, high life expectancy). 
In the breakdown by \texttt{Region} (Figure~\ref{fig:set_intent} right), she finds countries in Sub-Saharan Africa (yellow points) tend to be on the bottom right, with lower life expectancy and higher inequality. 
\begin{figure}[h!]
    \includegraphics[width=\linewidth]{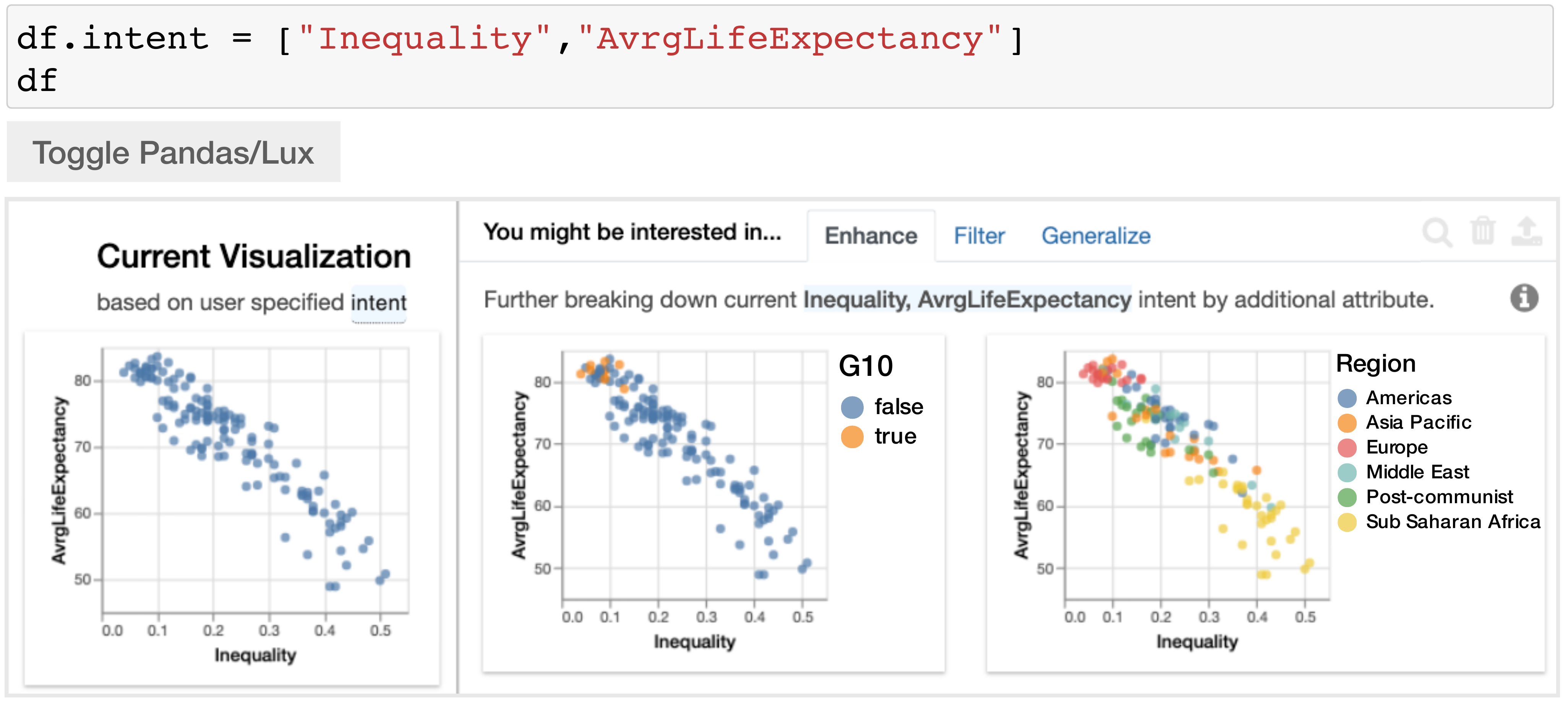}
    \caption{Alice sets the intent based on the attribute \texttt{AvrgLifeExpectancy} and \texttt{Inequality}, and \lux displays visualizations that are related to the intent.}
    \label{fig:set_intent}
\end{figure}
\stitle{Seamless integration with cleaning and transformation.} 
Alice is interested in how a country's 
development indicators relate to their early COVID-19 response
as of March 11, 2020. 
To investigate this, she imports a new dataset that characterizes how strict a country's response is, via \texttt{stringency}~\cite{oxford_covid}, a number from 0-100, with 100 being the highest level of responses.  As shown in Figure~\ref{fig:etl}, (I) Alice loads and joins the newly-cleaned dataframe with the earlier HPI dataset. 
(II) When she sets the intent on \texttt{stringency}, she finds that China and Italy have the strictest measures (dark blue on map Figure~\ref{fig:etl} center). She also learns that the histogram of \texttt{stringency} is heavily right-skewed (Figure~\ref{fig:etl} left), revealing how many countries had low levels of early pandemic response. (III) To better discern country-level differences, Alice bins \texttt{stringency} values into a binary indicator, \texttt{stringency\_level}, showing whether a country had \code{Low} or \code{High} levels of early response.
\begin{figure}[h!]
    \includegraphics[width=\linewidth]{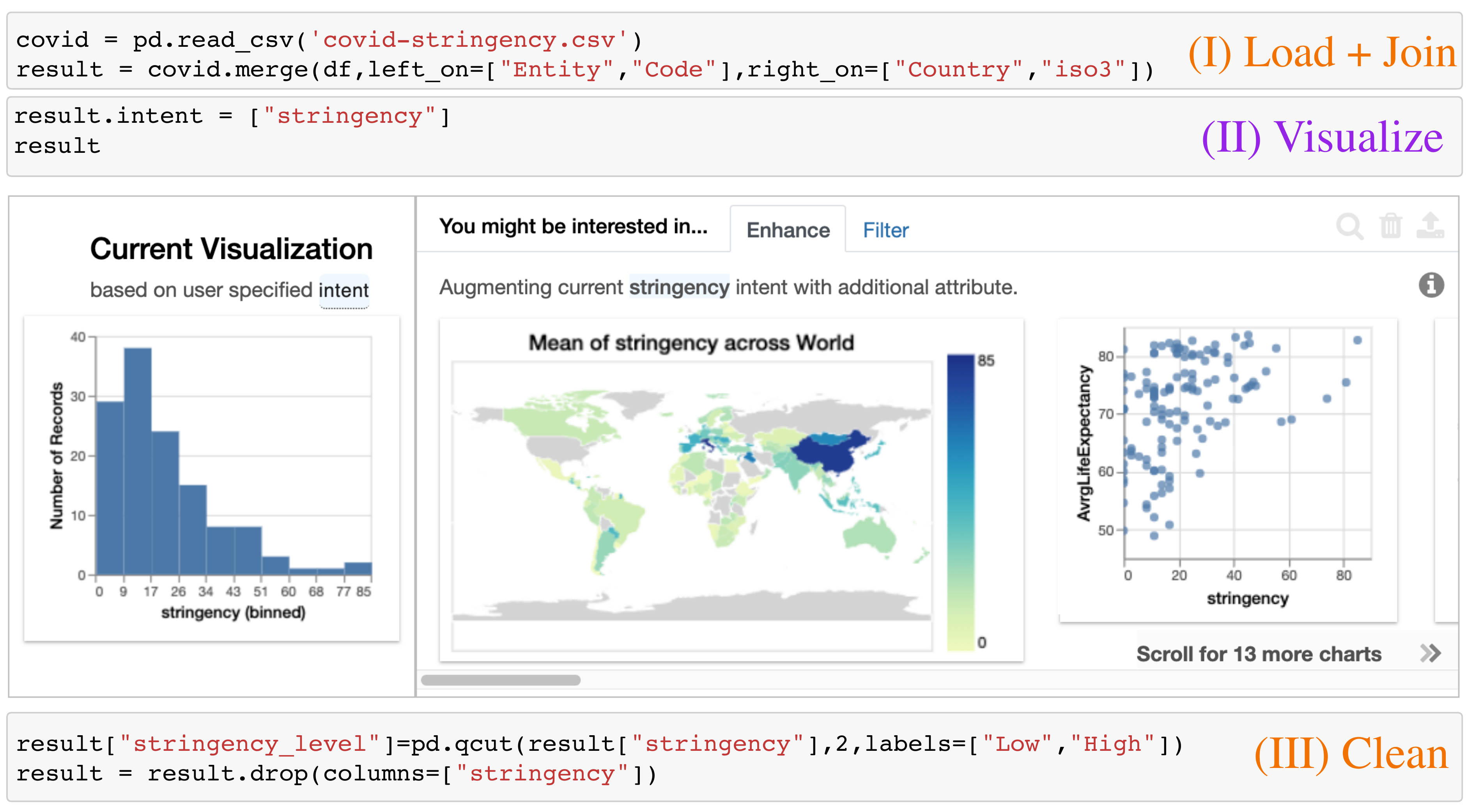}
    \caption{Tabular operations (orange,  steps I \& III) to load, clean, and transform the data, while visualizing with \lux (purple, step II).}
    \label{fig:etl}
\end{figure}
With the modified dataframe, Alice revisits the negative correlation she observed previously by setting the intent as 
average life expectancy and inequality again. The resulting recommendations are similar to Figure~\ref{fig:set_intent}, with one additional visualization showing the breakdown by \texttt{stringency\_level} (Figure~\ref{fig:insight} right). Alice finds a strong separation showing how stricter countries (blue) corresponded to countries with higher life expectancy and lower levels of inequality. This visualization indicates that these countries have a more well-developed public health infrastructure that promoted the early pandemic response. However, we observe three outliers (red arrow on Figure~\ref{fig:insight} right) that seem to defy this trend. When she filters the dataframe to learn more about these countries (Figure~\ref{fig:insight} left), she finds that these correspond to Afghanistan, Pakistan, and Rwanda---countries that were praised for their early pandemic response despite limited resources~\cite{afghanistan_covid,pakistan_covid,rwanda_covid}.
She clicks on the visualization in the \lux widget and the \includegraphics[height=1em,trim=0 10 0 0]{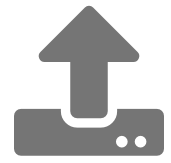} button to export the visualization from the widget to a \vis object. Alice can access the exported \vis via the \texttt{df.exported} property and print it as code, following which she can tweak the plotting style before sharing Figure~\ref{fig:insight} (right) with her colleagues. 

\begin{figure}[h!]
    \includegraphics[width=\linewidth]{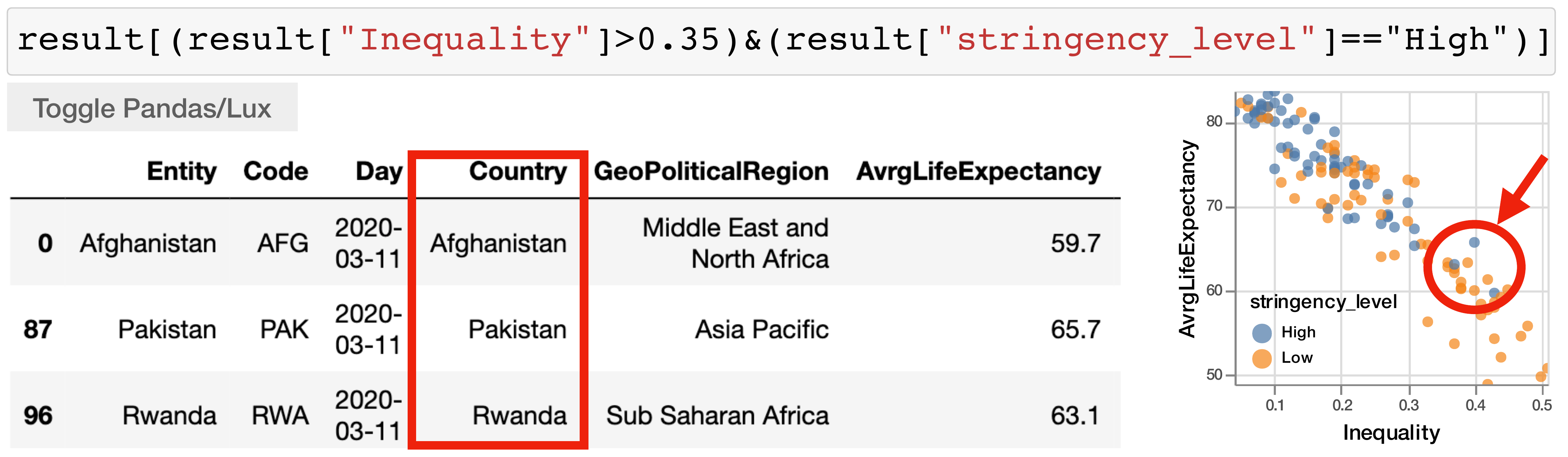}
    \caption{The scatterplot shows a separation between countries with high and low stringency in their COVID response. By filtering the dataframe (left), we see that Afghanistan, Pakistan, and Rwanda correspond to the three outliers (red boxed) that defies the trend.\label{fig:insight}}
\end{figure}

\par Overall, this example demonstrates the value of always-on visualization support within a dataframe workflow:
the tight integration between \lux and dataframes enabled Alice to seamlessly perform \change{data cleaning via a familiar API} and notebook environment. 


\section{Interacting with Dataframes}\label{sec:framework}

In this section, we propose a novel always-on framework for visual interaction with dataframes as outlined in Figure~\ref{fig:framework}.
The example workflow illustrated the many flexible ways 
users interact with a 
dataframe to achieve their 
analytical goal. 
Here, we summarize
these ways and contrast it to 
existing visualization specification
approaches in dataframe workflows.

As shown in Figure~\ref{fig:framework}, in both existing dataframe workflows (a) and the always-on framework (b), users can work with the dataframe API and see the table view by default (grey). For creating visualizations in an existing workflow, 
as shown in Figure~\ref{fig:framework}a, users would typically need to explicitly 
write visualization specification code in a language such as \texttt{matplotlib} or \texttt{altair} (orange) to create individual visualizations (blue). In our always-on framework, as shown in Figure~\ref{fig:framework}b,
users further inspect a dashboard of recommended visualizations,
as part of a multi-tiered framework (blue), all of which is driven by a user- or system-specified intent (orange), described below.




\noindent {\bf Intent}: Users can indicate aspects 
of the dataframe that they are interested 
in via a lightweight intent specification (\cref{sec:intent}). 
The intent drives the visualizations, 
actions, and dashboard. 
In the example, Alice indicated 
that she wants to learn  
about \texttt{AvrgLifeExpectancy} and \texttt{Inequality}; 
\lux displayed visualizations 
related to these variables. 
Unlike existing visualization libraries, 
intent can also be system-specified---meaning 
that the visual display will be always-on, 
even if the user does not explicitly specify intent.
We now describe the different layers in our always-on framework, following the notation in Figure~\ref{fig:framework}.

\noindent {\bf \circled{A} Visualization}: Visualizations 
are created by applying the intent to a given dataframe.
Each visualization, i.e., \vis, is an intent 
operating on a specific dataframe instance; 
a collection of visualizations is known as a \vislist. 

\noindent {\bf \circled{B} Actions}: Each action is 
an ordered collection of visualizations (\vislist),
e.g., the \code{Correlation} action plots pairwise relationships 
ranked by Pearson's correlation. 


\noindent {\bf \circled{C} Dashboard}: 
A dashboard is composed of one or more actions that may be relevant to the user.


Users can either make changes to the dataframe 
or the intent in order to fulfill different analytical needs. 
Dataframe operations are exact, 
leveraging the expressiveness of the dataframe API. 
On the other hand, the intent is a high-level 
specification of user interest, either explicitly
provided by the user or system-inferred, 
steering 
\lux's recommendations. 

By working with both the intent and dataframe API, 
\lux supports a flexible and intuitive 
experience for interacting with data.
Next, we describe the intent grammar 
that underlies the always-on framework of \lux. 

\begin{figure}[h!]
    \centering
    \includegraphics[width=\linewidth]{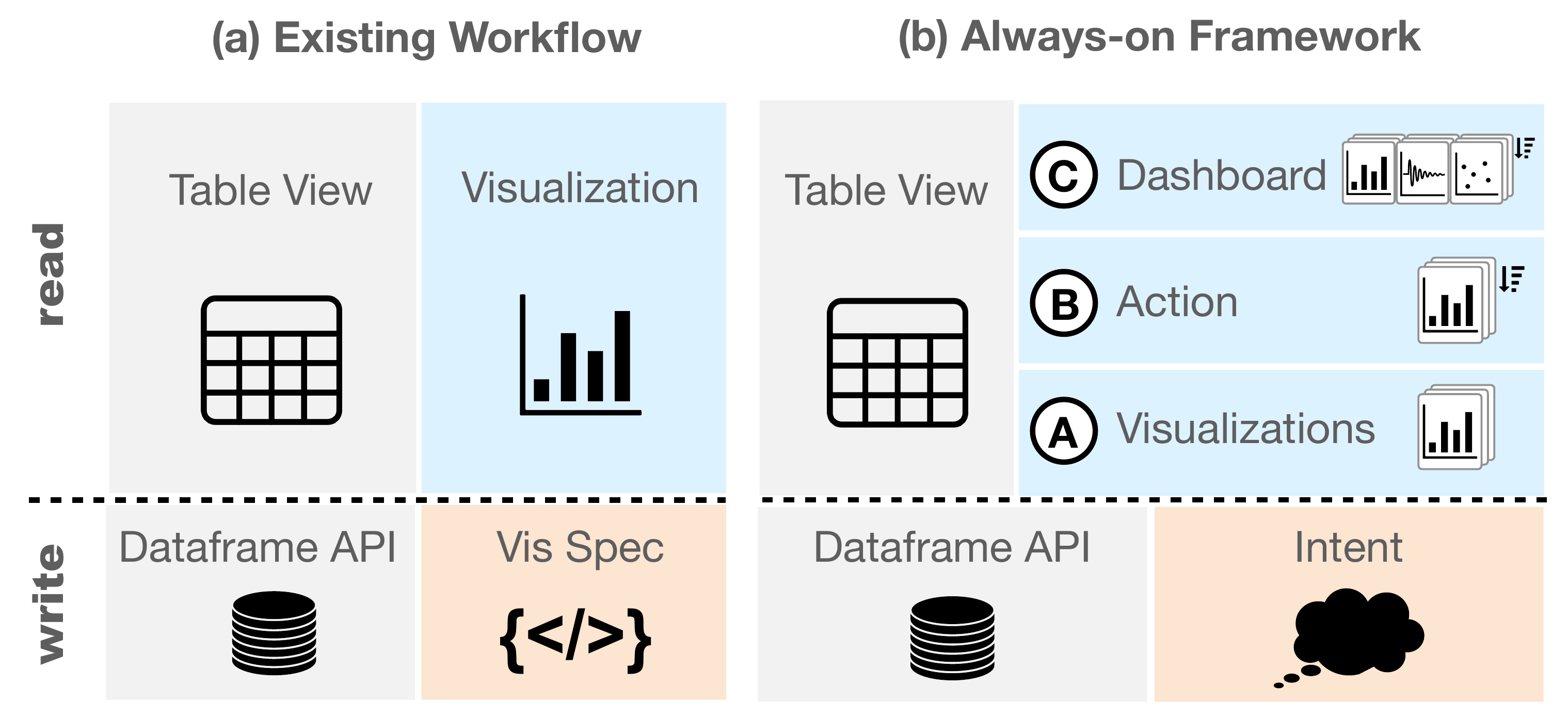}
    \caption{\change{Conceptual framework for dataframe interaction. Users can make changes to anything below the dotted line (\textbf{write}), elements displayed to the user are shown above the dotted line (\textbf{read}). (a) In existing workflows, users write visualization specification code to create one or more visualizations. (b) In \lux's always-on framework, users can optionally make changes to the intent, which steers the recommended visualizations (Visualizations, Action, Dashboard).}}
    \label{fig:framework}
\end{figure}

\section{Intent Language Formalization}\label{sec:intent}
The {\em intent language} is a lightweight, succinct means
for users to declaratively specify their high-level interests.
In this section, we 
introduce
this language and its underlying grammar, and how it differs from existing approaches. 

\subsection{Intent Grammar}\label{sec:set_intent}
The {\em intent} grammar describes what the user is interested
in within a dataframe. 
The intent is composed of one or more {\em clauses},
each of which is either an {\em axis} or a {\em filter}
of interest.
{\small
\begin{equation}
\begin{aligned}
    \langle Intent\rangle & \rightarrow \langle Clause\rangle^+ \\
    \langle Clause\rangle & \rightarrow \langle Axis \rangle \ \ | \ \  \langle Filter \rangle 
\end{aligned}
\end{equation}
}
An {\em axis} defines one or more attribute(s),
mapped appropriately to a specific encoding or channel
of the corresponding visualizations.
{\small
\begin{equation}\label{eq:axis}
    \langle Axis \rangle \rightarrow \langle attribute \rangle^* \langle channel \rangle  \langle aggregation \rangle  \langle bin\_size \rangle 
\end{equation}        
}
For the axis, apart 
from the mandatory attribute(s), specified under $\langle attribute \rangle$, the remaining properties
are optional---and can be automatically inferred.

{\em Filters} define a subset of data that the user is interested
in. To specify a filter, the attribute being filtered,
the operation, and the value, are required.
{\small
\begin{equation}\label{eq:filter}
    \langle Filter \rangle \rightarrow \langle attribute\rangle \ \  [=><\leq\geq\neq] \ \ \langle value\rangle
\end{equation}
}
Consider the simple case when $\langle attribute\rangle$ refers
to a single attribute and $\langle value \rangle$ refers
to a single value in Equations~\ref{eq:axis} and \ref{eq:filter};
then, an intent with multiple 
clauses (axis or filter) represents
a user preference to see 
each of the axis attributes visualized, 
for the subset of data corresponding
to the conjunction of the filters.

In the more general case, 
$\langle attribute\rangle$ can correspond
to a union of attributes, or a special wildcard value {\bf \circled{?}} (with an optional constraint to define the subset of attributes),
while the $\langle value \rangle$ can refer
to a union of values, or a special wildcard value {\bf \circled{?}}.
{\small
\begin{align}
        \langle attribute \rangle 
        &\rightarrow \textrm{attribute}\, \cup \, \langle attribute \rangle^* \  |  \ 
    \textrm{{\bf \circled{?}}} \ \langle constraint  \\ 
        \langle value \rangle 
        &\rightarrow \textrm{value}\, \cup \, \langle value \rangle^* \ \ \ | \ \  \textrm{{\bf \circled{?}}}
\end{align}
}
The use of unions in either case (as well as {\bf \circled{?}} 
which implicitly is a union of all alternatives)
admits a disjunction of options for 
the axis or filter clause.
If there are $n_i \geq 1$ alternatives for the $i^{th}$
clause, we can construct a collection of 
$n_1 \times n_2 \times \ldots \times n_k$
visualizations by taking the cross-product
of alternatives per clause.
\tr{Constructing a 
collection of visualizations
via partial specification of this sort has been 
explored in ZQL~\cite{Siddiqui2017} 
and CompassQL~\cite{wongsuphasawat2016towards}.}

\subsection{Specifying Intent}

\tr{The aforementioned grammar is decoupled from our specific implementation,
which uses syntactic sugar for expressing the intent
in a convenient Python-based API.} 
Users can specify an intent indicating their analysis 
interests or create
desired visualizations by applying the intent
to a specific dataframe.  We note that while the focus here is describing user-specified intent, the same intent language is used by the system for generating recommendations as will be described in Section~\ref{sec:rec}.

\subsubsection{Attaching an Intent to a Dataframe}\label{sec:syntax-intent}
Building on the grammar described above,
within \lux, a \clause can specify one or more columns (i.e., \axis) 
or rows (i.e., \filter) of interest.
\query{Q1} To set \code{Age} and \code{Education} as columns of interest
for a given dataframe \code{df},
one can state:
{\small \begin{verbatim}
    axis1 = lux.Clause(attribute="Age") 
    axis2 = lux.Clause(attribute="Education") 
    df.intent = [axis1,axis2]
\end{verbatim}
}
Or one can also use the equivalent shortcut:
{\small
\begin{verbatim}
    df.intent = ["Age", "Education"]
\end{verbatim}}
Once the intent is set, whenever \code{df}
is printed, the \lux widget will use the 
intent to determine what visualizations
to show to the user. 
Here, \lux would display 
visualizations 
related to attributes 
\texttt{Age} and \texttt{Education} from \code{df}.

We can compose \axis and \filter together, as follows.
\query{Q2} Explore the \texttt{Age}s for employees in the Sales \texttt{Department}. 
{\small
\begin{verbatim}
    axis = "Age"
    filter = "Department=Sales"
    df.intent = [axis, filter]
\end{verbatim}
}
\noindent Based on the specified intent, \lux not only shows the Age distribution filtered to the Sales department, but also displays a set of related visualizations, such as visualizations involving one additional attribute or one additional filter. 


In the following, we will showcase 
the \lux intent syntax as part of \vis
and \vislist, but the syntax can also
be used to simply set intent as 
in \code{df.intent} above.

\subsubsection{Constructing Visualizations Directly via Intent}\label{sec:syntax-vis}
Instead of attaching an intent to a
dataframe, one can use the \vis and \vislist keyword
to directly generate specific visualizations.
\query{Q3} Compare average \texttt{Age} across different \texttt{Education} levels.
{\small
\begin{verbatim}
    axis1 = lux.Clause(attribute="Age") 
    axis2 = lux.Clause(attribute="Education") 
    Vis([axis1,axis2],df)
\end{verbatim}
}
Query 3 is similar to Query 1, 
except that the intent is immediately
applied to the dataframe \code{df}
to create a visualization via \vis, 
rather than changing the intent associated with
the dataframe (to be used when the dataframe is eventually printed).
Given that the intent involves 
one measure (\texttt{Age}) and 
one dimension (\texttt{Education}), 
\lux will display a bar chart. 
By default, average is the function used for aggregation.

Aggregation is one of
three optional properties for \axis (Equation~\ref{eq:axis});
others are channel and binning.
If any of these are explicitly specified, 
they override \lux's defaults,
as in the following query.
\query{Q4} Compare the variance of \texttt{MonthlyIncome} 
based on employee \texttt{Attrition}.
{\small
\begin{verbatim}
  axis1 = lux.Clause("MonthlyIncome", aggregation=numpy.var)
  axis2 = "Attrition"
  Vis([axis1,axis2],df)
\end{verbatim}
}
\par To generate multiple visualizations, one could
either set \code{df.intent} as in Section~\ref{sec:syntax-intent},
which would generate a collection of visualizations related
to the intent, or specify intent 
as an input to a \vislist,
as in the following query.
\query{Q5} Show how factors related 
to the rate of compensation differ 
for employees with different \texttt{EducationField}s. 
{\small
\begin{verbatim}
    rates = ["HourlyRate","DailyRate","MonthlyRate"]
    VisList(["EducationField",rates],df)
\end{verbatim}
}
Here, there is one \vis corresponding to \code{EducationField}
combined with each of \code{HourlyRate}, \code{DailyRate},
and \code{MonthlyRate}.
The wildcard character {\bf \circled{?}},
when used as part of an \axis,
can be used to enumerate 
over \textit{all} attributes in a dataframe;
constaints may be used to restrict them to a certain type. 
\query{Q6} Browse through relationships between any two quantitative columns in the dataframe.
{\small
\begin{verbatim}
    any = lux.Clause("?",data_type = "quantitative")
    VisList([any, any],df)
\end{verbatim}
}
This \vislist 
corresponds to the search space for the \code{Correlation} action;
the \code{Correlation} action 
additionally ranks and sorts each \vis in the \vislist 
based on their Pearson's correlation score.

\filter values can also be specified as a list or via wildcards across all possible values for a fixed filter attribute. 
\query{Q7} Examine \texttt{Age} distributions across different \texttt{Countries}.
{\small
\begin{verbatim}
    VisList(["Age", "Country=?"],df)
\end{verbatim}
}
\npar The generated \vislist contains histograms of \code{Age}, 
one each for individuals where \code{Country} is \code{USA}, 
\code{Japan}, \code{Germany}, and so on.

\subsection{Rationale for Design Choices}\label{sec:intent_design_consideration}
Due to the heavy cognitive cost of writing glue code
to visualize their data~\cite{Wang_Viser2020,alspaugh2019}, 
users often opt to visualize
in the later stages of their workflow~\cite{Batch2018,Kandel2012Interview}.
Instead, our goal with \lux's intent language has been to 
support visualization to be used
throughout;
and for this, users should not have to expend too much effort in thinking 
about what and how to visualize. 
There are two key characteristics of our intent language
that support this quick and flexible programmatic specification,
described next. 

\stitle{Versatility:} 
Our intent language is {\em versatile} in that it 
serves both as a mechanism for steering recommendations (Q1-2)
and as a way of directly creating visualizations 
on top of dataframes (Q3-7). 
This is unlike existing specification approaches
whose sole focus is the creation of one or more visualizations. 
This versatility means that 
whenever users specify their intent, 
they are not committing to a pre-defined set of operations. 
Instead, the system leverages explicit user input 
(in the form of intent),
as well as implicit signals
to determine what to display to users.

Consider Q2,
which demonstrates the versatility
of intent beyond the specification of a single
visualization. 
Here, the user simply specifies the 
data-specific aspects they are interested
in, i.e., the attribute \texttt{Age},
and the \texttt{Sales Department} filter;
these are used as cues by \lux to generate
visualizations,
including those that wouldn't ordinarily be picked
if we were using a conventional visualization specification
framework (such as those with a different filter).
This versatility makes it easy for users
to communicate their analysis intent even when
they do not have a specific visualization in mind.

\stitle{Convenience:} 
Our intent language 
only requires specification of
data-oriented aspects,
while existing approaches also
require users to specify
visual encoding-oriented aspects
to generate visualizations.
Our minimalistic language design 
is intended to alleviate the common challenge 
in exploratory analysis 
where users struggle 
to translate their high-level 
data questions to exact visualization specifications~\cite{Grammel2010}. 
\lux supports convenient specification shorthands and defaults 
and automatically infers the necessary details to transform user-specified intent 
into complete specifications. 

As shown in Q3-7, where
the target is one or more specific visualizations,
\lux enables users to visualize their data 
with only a single line of code, 
effectively lowering the barrier to visual exploration. 
In Figure~\ref{fig:comparison}, we outline the code 
required to create a single visualization 
based on Query 3, and compare the key differences in the required specification
across various languages, including Draco~\cite{moritz2019formalizing}, matplotlib~\cite{matplotlib}, and Vega-Lite~\cite{satyanarayan2017vega}. 
Other languages often
require users to specify the field type, 
channel, and marks, while \lux can reason over underspecified intent. 
This reduces the effort required on the part of the user.

\begin{figure}[ht!]
    \centering
    \includegraphics[width=\linewidth]{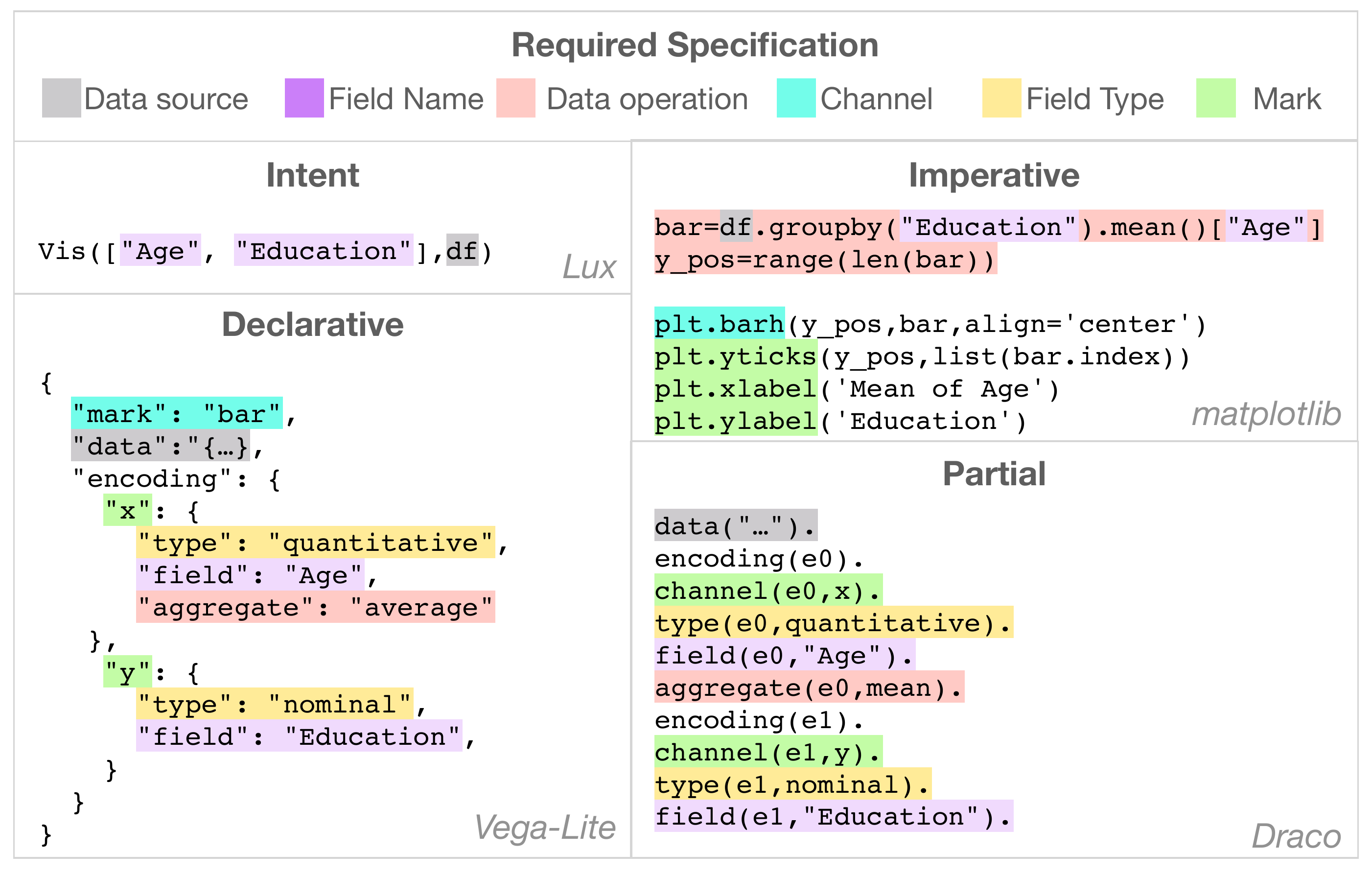}
    \caption{\change{Comparison between the level of specification required from \lux versus other existing approaches for Query 3.}}
    \label{fig:comparison}
\end{figure}
\section{\change{Dataframe} Recommendations}\label{sec:rec}
In the previous section, 
we have seen how users
can either attach an intent
to a dataframe, or this intent
can be programmatically
generated as part of
\lux's recommendations.
We discuss the latter in this section.
In \lux,
an \textit{action} 
describes 
a ranked list of visualization recommendations based on a predefined search space. 
\lux supports four major classes
of actions\tr{, as summarized in Table~\ref{table:action_tbl}}.
Metadata- and intent-based ones
\change{are akin to those used in past visualization
recommendation systems~\cite{frontier,hu2018dive,wongsuphasawat2017visualizing}---see
\tr{Lee et al.~\cite{frontier}} \papertext{our technical report~\cite{lux-tr}} for details. 
\change{As described in Section~\ref{sec:relatedwork}, most existing VisRecs are situated in GUI-based charting tools; our key novelty is that \lux is one of the first visualization recommendation systems that is designed to fit into a programmatic dataframe workflow. Specifically, here,} we introduce two novel classes of recommendations specific to dataframe-based 
workflows}, based on dataframe structure and history.
\tr{
    \begin{table}[h!]
        \includegraphics[width=\linewidth]{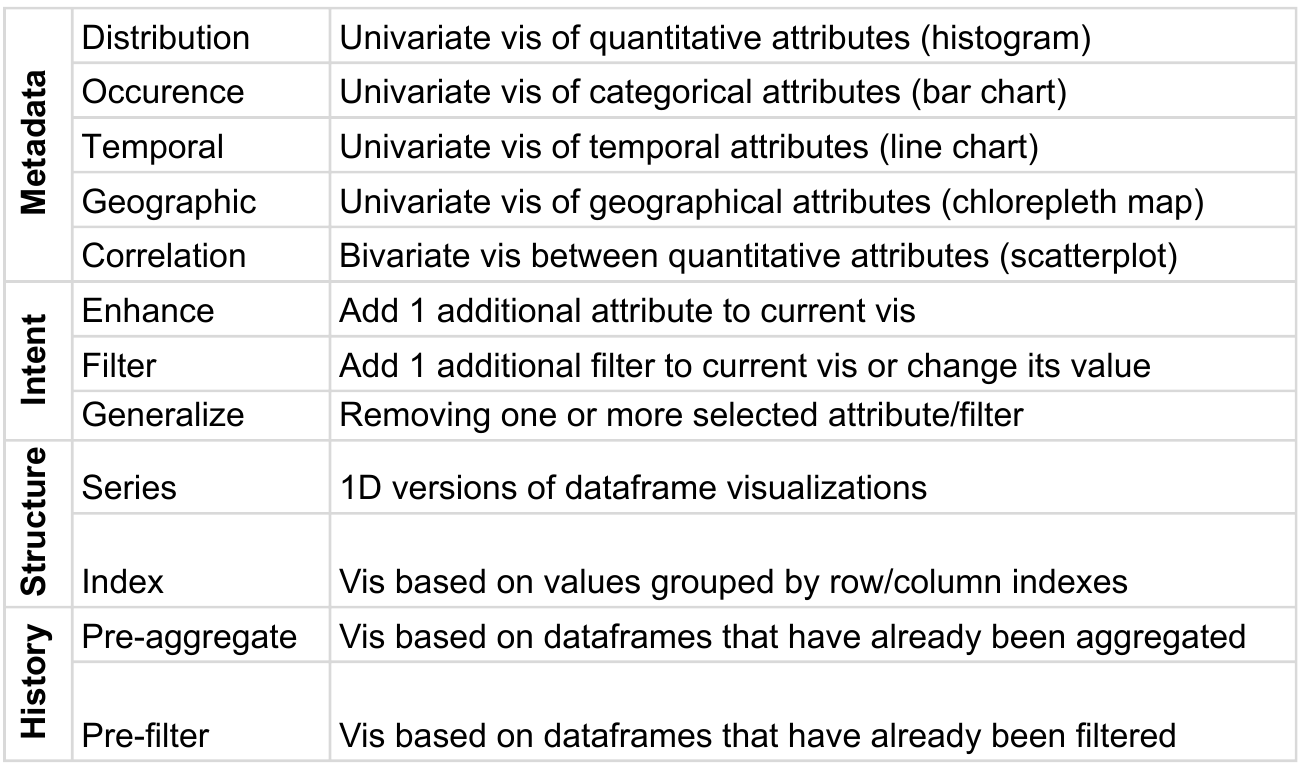}
        \caption{Different types of default recommendations in \lux}
        \label{table:action_tbl}
    \end{table}
    \stitle{Metadata-based Recommendations.}
    \lux maintains dataframe metadata, 
    including attribute-level statistics
    such as min/max and cardinality
    to determine the semantic 
    data type of each column and 
    to automatically populate visualization
    settings.
    For example, based on data type,
    \lux can generate univariate and bivariate
    overviews. In Figure~\ref{fig:print_df}, 
    Distribution, Occurrence, Temporal, and Geographical actions
    provide univariate overviews of columns,
    while the Correlation action provides bivariate overviews of all
    possible pairs of quantitative attributes,
    ranked based on Pearson's correlation.
    \tr{Metadata-based recommendations have been used extensively in past visualization recommendation systems~\cite{hu2018dive,wongsuphasawat2017visualizing}.}

    \stitle{Intent-based Recommendations.}
    \lux displays recommendations based
    on the user-specified intent. 
    On printing the dataframe,
    \lux displays a visualization based
    on the user-specified intent
    as in Figure~\ref{fig:set_intent},
    as the \textit{Current Visualization}.
    In addition, \lux provides
    recommendations based on valuable
    next analysis steps starting from that visualization.
    For example, the \code{Enhance} action
    recommends visualizations formed by adding 
    an additional attribute 
    to the current visualization. 
}

\stitle{Structure-based recommendations.}
\change{Data scientists often reshape their dataframes 
in ways that are more amenable to downstream
analysis, modeling, or presentation. One of our key insights is that the dataframe ``structure'' reveals strong signals for what
the users subsequently choose to visualize, thus providing implicit information on what recommendations to display
automatically by \lux.}

\emtitle{Index-based visualizations:} 
Dataframe indexes provide
a natural way to order and label 
dataframe rows and columns.
Indexes are
typically created as a result
of grouping and aggregation through operations 
such as \texttt{groupby}, \texttt{pivot}, \texttt{crosstab}. 
For any \emph{pre-ag\-gregated} 
dataframe 
(i.e., dataframes resulting 
from an aggregation operation), 
\lux creates visualizations 
by grouping the values 
row or column-wise.
For example, Figure~\ref{fig:indexgroup} displays 
the result of a pivot operation, 
where each row is visualized as a time series line chart. 
\tr{\lux currently only supports single-level indexes, visualization of multi-level indexes is a potential direction for future work.}

\emtitle{Series visualizations:} 
Series are 
dataframes with a single column. 
\lux leverages the 
same dataframe visualization mechanism for Series,
displaying univariate, metadata-based visualization, 
such as a bar chart for categorical
and histogram for quantitative Series.
\tr{\noindent By visualizing dataframe structure, 
    \lux provides a natural and intuitive representation 
    of dataframes and their derivative products. 
    These visual representations 
    can be extended to other dataframe-derived 
    structures (e.g., \texttt{GroupBy}, \texttt{Offset}, or \texttt{Interval}) 
    to help novices learn, debug, 
    and validate complex dataframe operations.
}
\begin{figure}[h!]
    \centering
    \includegraphics[width=0.9\linewidth]{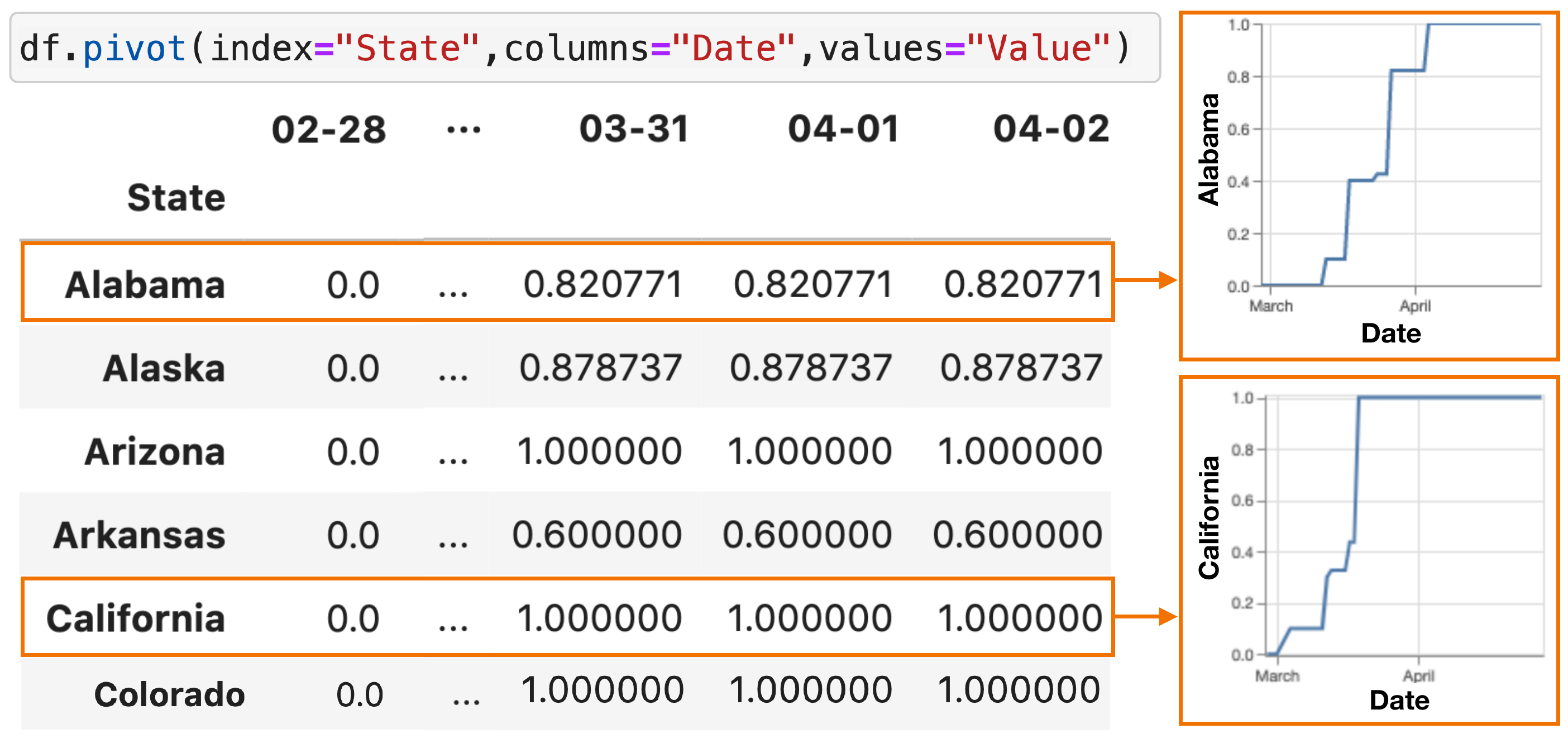}
    \caption{Row-wise index visualization displaying the normalized percentage of COVID-19 cases across different \texttt{State}s.}
    \label{fig:indexgroup}
\end{figure}

\stitle{History-based recommendations.}
Apart from dataframe structure, another source of implicit information from the dataframe is the historical set of operations performed by users.
\tr{For example,
if the user cleaned up a particular column
and renames it,
it is likely that they would want to 
visualize the same column soon thereafter.}
\lux displays history-based recommendations
based on whether the dataframe
has been filtered or aggregated
in its recent history.
For example, when a filtering-based operation 
leads to a small dataframe 
(such as when a \texttt{head} or \texttt{tail} is performed), 
\lux visualizes the previous unfiltered dataframe 
since there are too few tuples 
for generating recommendations 
in the filtered dataframe.
\lux also uses history to determine
if an aggregation has been performed,
helping identify the structure-based recommendations
described earlier.

\tr{
    To collect this history, 
    since \lux acts as a wrapper around \pandas
    (described in the next section), 
    we instrument each dataframe function and 
    track each one with minimal overhead and store it 
    as part of the dataframe, instead of requiring program analysis, 
    which is prone to false positives~\cite{yan2020auto-suggest}. 
    Given that new dataframes or intermediate 
    objects (e.g., \texttt{GroupBy}, \texttt{Series}) 
    are often created when the user performs an operation, 
    \lux propagates the history 
    over to derived objects so that 
    the history is not lost.
    A key challenge for leveraging dataframe history to infer better recommendations would be around surfacing the inferred implicit intent in a way that is interpretable and explains resulting recommendations choices.
}


\section{System Description}\label{sec:system}

\lux employs a client-server model, 
leveraging computational notebooks 
as a frontend client. \tr{\lux currently supports Jupyter Notebooks, 
Jupyter Lab, 
Jupyter Hub, 
Microsoft Visual Studio Code, 
and Google Colab. The \texttt{ipywidgets} library 
is used for rendering an 
interactive HTML widget as the cell output.} 
Once users import \lux,
they can interact with a \luxdf
instead of a regular \pandas dataframe.
\luxdf acts as a wrapper
around \pandas, and supports all
existing \pandas operations,
while storing additional information,
such as the intent, metadata, structure,
and history, for generating visual
recommendations.
As shown in Figure~\ref{fig:architecture}, 
the server side logic is largely separated 
into two distinct layers: 
1) the \textit{intent processing} layer 
is responsible for processing 
intent into executable instructions, 
and 
2) the \textit{recommendation} layer 
is responsible for generating 
the displayed visualizations. 
To generate the visualization recommendations, 
as well as compute metadata that is used 
in various stages,
the execution engine 
performs the required 
data processing and optimization, either as a series of dataframe operations 
in \pandas or equivalently in SQL queries
in relational databases (\cref{sec:compute}). Finally, the system design is modular 
and extensible so that alternatives 
can be swapped in at different layers, e.g., Altair and Matplotlib
visualization rendering libraries.

\begin{figure}[ht]
    \centering
    \includegraphics[width=0.85\linewidth]{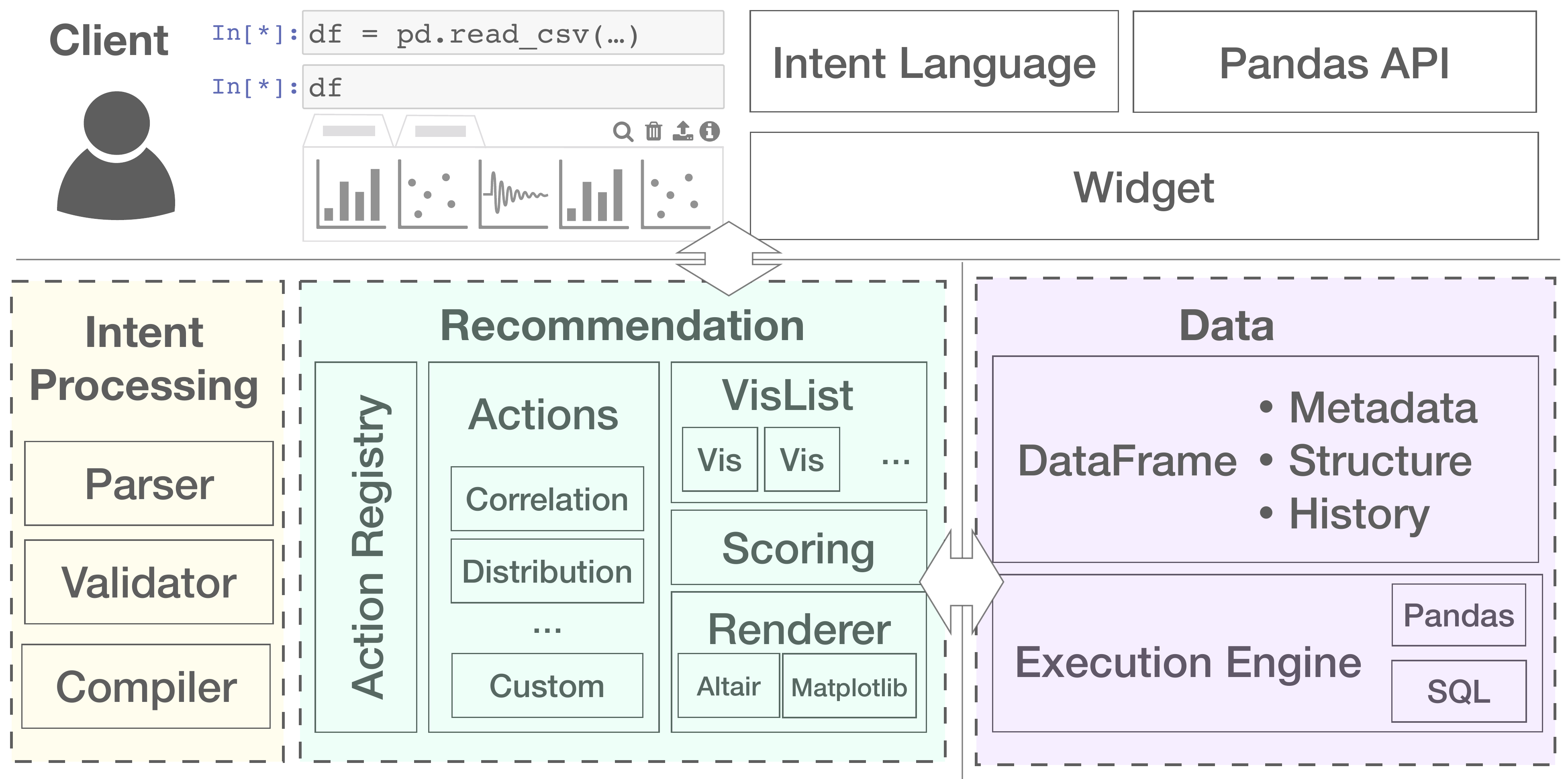}
    \caption{System architecture for \lux}
    \label{fig:architecture}
\end{figure}
\tr{
    \subsection{Intent Processing}\label{sec:intent_process}

    Here, we discuss how \lux processes user intent 
    to automatically infer missing details 
    and determine appropriate visualization mappings.
    The intent processing layer parses, validates, and 
    compiles the user's underspecified intent 
    into complete specifications. 

    \subsubsection{Parser and Validator}\label{sec:parse-validate}

    In Section~\ref{sec:intent}, 
    we saw how \axis and \filter can be
    be used to compose \clauses;
    the parser parses the user-inputted strings
    into an internal \clause representation.
    Subsequently, the validator
    checks for any inconsistencies 
    between user-specified \clauses 
    and the dataframe content. 
    To do so, it leverages the dataframe's 
    pre-computed metadata 
    to verify the input intent.
    If the user’s input does not align 
    with the data present in the dataframe, 
    the validator provides early warnings 
    and suggests corrections to the input intent. 

    \subsubsection{Compiler}\label{sec:compile}
    During intent specification,
    users have the ability to omit
    certain optional details,
    making them {\em partial specifications}.
    Users also implicitly construct
    a collection of visualizations 
    by using a union or wildcard character
    for \axis or \filter. 
    Post validation, the compiler 
    expands the \clauses
    into multiple visualizations 
    and adds in defaults for the omitted
    details, making the \clauses \ {\em complete}.
    This transformation is performed in three steps. 
    \stitle{1) Expand:} 
    If the input intent implicitly 
    encodes multiple visualizations,
    the compiler ``unrolls''
    these visualizations into individual
    \vis objects as a cross-product
    of the specified \clauses,
    leading to a \vislist
    containing the resulting 
    visualization specifications.

    \stitle{2) Lookup:} For each  
    \vis in the \vislist, 
    \lux populates the omitted details  
    using the dataframe's pre-computed metadata. 
    The compiler also removes any 
    invalid visualizations generated 
    that are either not supported in \lux 
    or use ineffective encodings.

    \stitle{3) Infer:} Finally, 
    \lux infers the visualization encodings, 
    including the marks, channels, and 
    transforms (sort, aggregation, binning) 
    required for generating the visualizations. 
    The compiler implements rule-based heuristics 
    drawn from best practices in design~\cite{Few2012,Mackinlay2007}. 

    \smallskip
    \noindent 
    After intent processing, 
    \lux can now use the complete intent specification 
    to either generate a \vis directly or 
    generate a set of appropriate recommendations (described next).

    \subsection{Recommendation Generation}\label{sec:rec_gen}

    As described in the framework in Figure~\ref{fig:framework}, 
    actions organize collections of views 
    into recommendations displayed to the users. 
    The \textit{action registry} in \lux 
    keeps tracks of a list of possible actions 
    that could be applicable for 
    generating recommendations 
    at any point in the analysis. 
    On initialization, 
    \lux registers a set of default actions 
    (described in Section~\ref{sec:rec}) 
    applied to all dataframes. 
    Users can also register their 
    own custom actions programmatically 
    by writing a Python-based UDF. 
    The UDF generates a \vislist of possible visualizations 
    and optionally scores and ranks each \vis. 
    The custom action is ``triggered'' whenever 
    the dataframe satisfies the user-specified 
    condition on when the action is applicable;  
    \lux recommends visualizations based on the action.

}

\section{Execution and Optimization}\label{sec:compute}

\papertext{
    \lux's execution engine performs two major tasks: 1) 
    compute the metadata required to generate candidate visualizations, 
    and 2) extract the actual data for each visualization.
    \stitle{Metadata Computation:}
    The metadata computed includes 
    attribute-level statistics
    and data types. 
    The statistics include
    the list of unique values, 
    cardinality, 
    and min/max values for each attribute. 

    \stitle{Visualization Processing:}
    After the user or system-specified 
    intent has been transformed 
    into one or more visualizations
    with a complete specification, 
    the execution engine translates 
    each visualization to queries responsible for processing 
    the data required for the visualizations. 
    The engine applies any filters and projections on the data before performing 
    different visualization-specific operations, such as binning and aggregation.  
}
\tr{
We now describe \lux's execution engine. 
    We first describe the 
    two major tasks performed
    by this execution engine.
    Then, we describe three optimizations
    aimed at speeding up these tasks.

    \subsection{Execution Engine}\label{sec:exec-engine}

    We now discuss how we
    computes metadata and visualizations.

    \stitle{Metadata Computation:}
    The metadata computed includes 
    attribute-level statistics
    and data types. 
    The statistics include
    the list of unique values, 
    cardinality, 
    and min/max of the attribute. 
    The unique values 
    \change{are} used to determine the 
    candidates generated 
    by a wildcard 
    for a filter on the column,
    or for validating
    filter input for the column,
    and for computing the cardinality. 
    The cardinality information 
    is used to determine the data type,
    while min/max is used for 
    determining the limits on 
    the visualization axes. 
    Next, the execution engine 
    infers the semantic data type 
    based on the internal data type 
    and cardinality information. 
    \lux supports nominal, 
    quantitative, geographic, 
    and temporal data types. 
    If the data type is misclassified, users can override the automatically-inferred data type.

    \stitle{Visualization Processing:}
    After the user or system-specified 
    intent has been transformed 
    into one or more \vis objects 
    with a complete specification, 
    the execution engine translates 
    each \vis to queries responsible for processing 
    the data required for the visualizations. 
    First, the engine applies any filters 
    and retrieves relevant attributes. 
    Next, the execution engine performs 
    different visualization-specific operations 
    depending on the mark type. 
    For example, to process the data 
    for a histogram, the engine bins an attribute into fixed-sized bins and performs a count aggregation for each bin. Table~\ref{tbl:vis_relational_processing} summarizes the relational operations that corresponds to processing different visualization types. 

    \subsection{Optimization}
    Next, we describe several optimizations aimed at minimizing the overhead incurred by \lux.
}
    Computing metadata and processing data for visualizations can be time consuming, 
    even for a moderately-sized dataframe. Therefore, we adapt optimizations from approximate query processing~\cite{GarofalakisG01, CormodeGHJ12}, early pruning~\cite{Vartak2015,kim2015rapid,fastmatch}, caching and reuse~\cite{GuptaM05, TangSEKF19}, and asynchronous computation~\cite{Xin2021,BendreWMCP19}, 
    to improve the interactivity of \lux.
\stitle{Intelligent workflow-based optimizations (\wflow):}
During an analysis session, users constantly modify and operate on dataframes, which means that the metadata and associated recommendations can change throughout a session, 
especially during reshaping and type-modifying 
operations. \tr{Figure~\ref{fig:workflow} shows an example of one such workflow. }Thus, unlike conventional 
visual analytics, 
where metadata can be computed 
upfront 
and stays fixed throughout, 
here, 
metadata needs to be constantly updated 
to ensure that recommendations 
are generated correctly. 
As a result, the computation 
associated with keeping the 
metadata ``fresh'' after each dataframe operation can be computationally expensive.
We propose two techniques to reduce this overhead: 1) lazily compute the metadata and recommendations only when users explicitly print dataframes; 
2) cache and reuse results 
later on in the session.

\par Since users often intersperse dataframe printing 
with several data\-frame operations, it is likely that the computed metadata and recommendations would be outdated before users see the results. As a result, we can delay computation and compute the metadata and recommendations only after the user has explicitly requested to print a dataframe. Each \luxdf keeps track of how fresh the metadata and recommendations are and expires them when an operation makes a change to the dataframe. In particular, we leverage \pandas's internal functions that are triggered when: 
\begin{denselist}
    \item the dataframe is modified inplace instead of returning a new dataframe, e.g., \texttt{df.dropna(inplace=True)}
    \item columns in the dataframe are updated, either through the bracket or dot notation, e.g., \texttt{df.Frac} or \texttt{df["val2"]=df["val"]/5}
    \item the row or column labels are changed, e.g., \texttt{df.rename(columns}\\\texttt{={"val":"value"})}
\end{denselist}
Additionally, recommendations are expired when the intent is modified. On printing the dataframe, \lux recomputes metadata and generates the recommendations accordingly. This lazy strategy ensures no overhead  on any non-print operations. \tr{Future work on more intelligent, fine-grained maintenance and expiration strategies can improve system performance (e.g., only refresh metadata and recommendation relevant to a specific column instead of entire dataframe for a single column update).}

\lux further memoizes the metadata and recommendations so that any subsequent prints to an unmodified dataframe do not require recomputation. 
Users frequently perform ``non-committal'' operations that do not make changes to the dataframe to be used in subsequent analyses, involving printing dataframes as intermediate results to facilitate quick experimentation and debugging.
\tr{As shown in cells labeled [3-5] in Figure~\ref{fig:workflow}, }
\papertext{For instance, }users may print a column, perform grouping and aggregation, or print descriptive summaries, all without modifying the dataframe. In this case, when the user revisits the original dataframe, the memoized recommendations are immediately accessible to them. 
\papertext{
    \par While lazy computation and caching and reuse are well-studied~\cite{GuptaM05,TangSEKF19,xin2019}, identifying common dataframe usage patterns and determining when and how to expire
    metadata and recommendations in a dataframe workflow are novel contributions.
}
\tr{
    \par Note that while lazy computation and caching and reuse are well-studied in the database literature~\cite{GuptaM05,TangSEKF19,xin2019}, identifying that lazy computation may be beneficial
    non-committal dataframe operations is a novel insight. 
    Similarly, recognizing that when users
    repeatedly print the same dataframe
    without modifying it, caching
    and reuse could be valuable is our novel contribution.
    Combined, these observations around common dataframe usage patterns inspired our novel approach for determining when and how to expire
    metadata and recommendations in a dataframe workflow.
}
\tr{
    \begin{figure}[h!]
        \centering
        \includegraphics[width=0.9\linewidth]{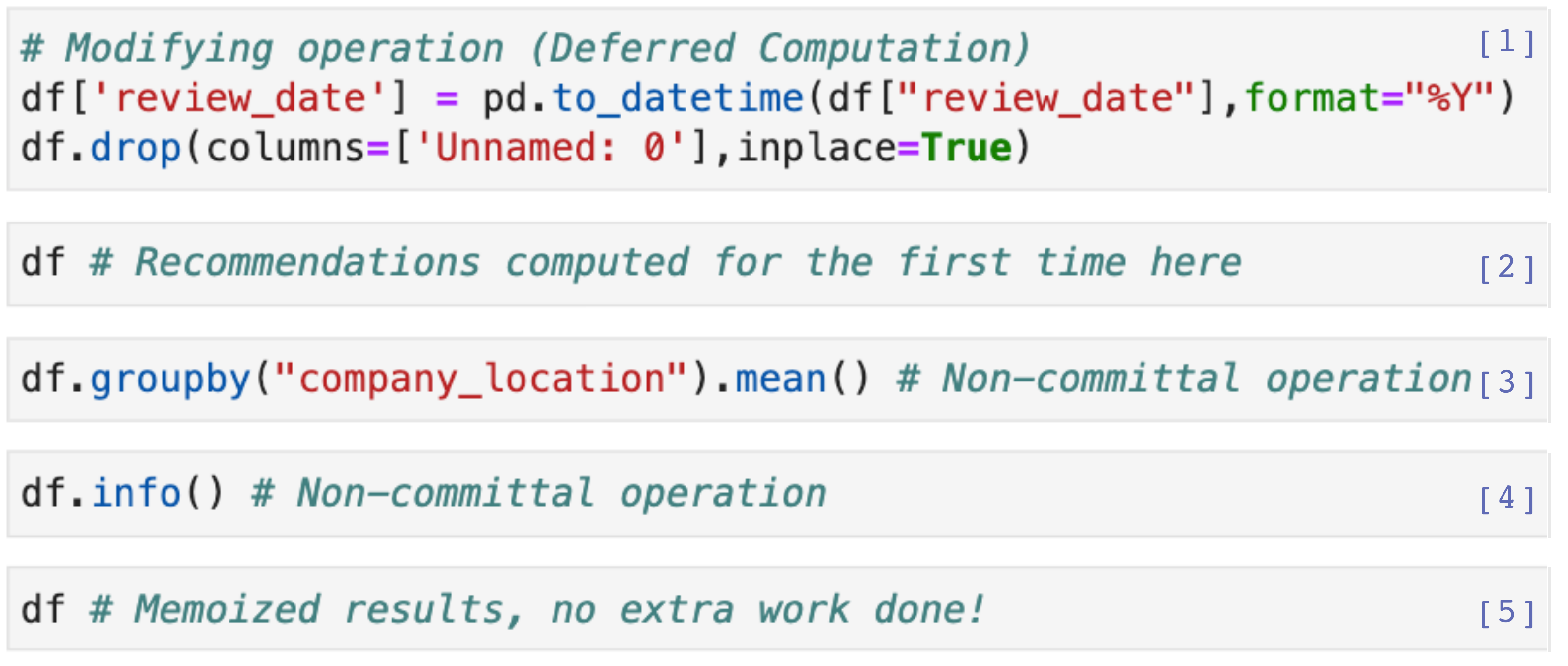}
        \caption{Example workflow demonstrating the applicability of \wflow optimizations.}
        \label{fig:workflow}
    \end{figure}
}
\stitle{Approximate, early pruning of search space (\prune):}
\change{As described in Section~\ref{sec:framework}, each visualization in an action is ranked based on a scoring function, computed based on the data associated with each visualization. Inspired by existing work in early pruning~\cite{Vartak2015,kim2015rapid,fastmatch}, \lux estimates the visualization score to speed up the retrieval of top-k visualizations for each action. We employ approximate query processing to reduce the cost by estimating the scores using sampled data. Specifically, \lux first performs a preliminary pass over the \vislist to approximate the score of each visualization and then proceeds to recompute the top-k selected visualizations in a second pass to process each of the displayed visualizations \textit{exactly}. Currently \lux leverages a cached sample of the dataframe to approximate visualization scores (e.g., for a dataframe with 1M rows, approximating correlation score by using only 30k rows), although other approximate query processing methods could be applied.}
\begin{table}[h!]
    \centering
    \caption{Table summarizing the relational operations performed for processing different visualizations. \tr{Primary operations that accounts for the bulk of the visualization processing costs are listed.}}
    \label{tbl:vis_relational_processing}
    \begin{tabular}{c|c}
    \textbf{Vis Type}         & \textbf{Relational Operation}                             \\ \hline
    Scatterplot      & Selection on 2 columns                    \\
    Color Scatterplot  & Selection on 3 columns                    \\
    Line/Bar         & Group-By Aggregation                      \\
    Colored Line/Bar & 2D Group-By Aggregation                   \\
    Histogram        & Bin + Count                           \\
    Heatmap          & 2D Bin + Count                        \\
    Color Heatmap  & 2D Bin + Count + Group-By Aggregation
    \end{tabular}
\end{table}
\par Given that the \prune optimization performs two passes over the \vislist (first pass for pruning, followed by an exact recomputation for the top-k), the additional recomputation cost incurred can be higher than doing a single pass over the \vislist. \change{For example, dataframes that are wide or contain high-cardinality attributes can often result in actions involving large visualization search spaces. }Therefore, this optimization should only be applied when the approximate savings are larger than the recomputation cost of the top k visualizations: $N\times t_{exact} \gg N \times t_{approx} + k \times t_{exact}$, where $N$ represents the number of candidate visualizations, $t_{exact}$ and $t_{approx}$ are the cost of computing the exact and approximate scores, respectively. \tr{Intuitively, in the ideal case where $t_{approx}$ is close to zero, $N$ needs to be at least greater than $k$ as a minimum requirement for the \prune optimization to provide meaningful savings.} \change{The cost of scoring a visualization, $t_{exact}$ and $t_{approx}$, is determined 
by the relational operations for extracting the required visualization data (as shown in Table~\ref{tbl:vis_relational_processing}).} 
\change{\par Here, while the use of approximate samples to rank and identify top-k visualizations
is not new~\cite{fastmatch,Vartak2015}, our use of approximation in conjunction with a cost model
to determine its potential interestingness is a novel application of the technique.}

\stitle{Cost-based scheduling of actions (\async):}
We find that users generally spend an average of 28 seconds\footnote{Based on 514 collected logs of Lux usage, the time spent on the initial \pandas table follows a long-tail distribution, with a median of 2.8 seconds and standard deviation of 183.4 seconds.} skimming through the \pandas table view before toggling to the \lux view. \change{Leveraging past work on asynchronous query execution~\cite{BendreWMCP19,Xin2021}, recommendation results can be streamed into the frontend widget as the computation for each action completes to ensure interactive responses,} without having to wait for all of the actions to finish rendering. After compiling the visualizations for each action, we estimate the cost of the action as the sum of the visualization costs in the \vislist, using the cost model \change{described in our technical report~\cite{lux-tr}}. This estimate is then used for scheduling the cheapest action to compute first, followed by computing the remaining in the background. In datasets where a few ``laggard'' actions dominate the overall recommendation generation (e.g., \code{Correlation} for a wide and highly quantitative dataset), the \async optimization provides users with early results and returns interactive control back to the user, instead of incurring a high wait time during their analysis session.

\change{\par The idea of exploiting asynchronous execution during user wait-time has been well-established~\cite{BendreWMCP19,Xin2021}, but our work is the first to apply this technique in a visualization recommendation context, by leveraging cost estimates to prioritize cheaper-to-compute visualizations. \change{Our cost model across different visualization types is an independent valuable contribution.}}

\section{Performance Evaluation}\label{sec:evaluation}
We evaluate \lux to measure its performance on large real-world datasets and notebook sessions\papertext{.}\tr{, along the following dimensions: 
\begin{denselist}
    \item RQ1: What is the overall performance of \lux? Can \lux achieve interactive latency during a typical dataframe workflow? 
    \item RQ2: What is the effect of the number of columns on \lux's performance?
    \item RQ3: How does the approximation-based \prune condition affect the quality of the recommendations relative to no approximation?
\end{denselist}
}We focus on evaluating the interactive latency in this section; we describe the usability evaluation in the following section. \change{Source code for experiments and analysis are available online}\footnote{https://github.com/lux-org/lux-benchmark}.

\tr{\subsection{Data and Methodology}}
\stitle{Data:} We use two real-world datasets to evaluate the performance of \lux. The \airbnb dataset~\cite{airbnb_notebook} contains 12 columns while the \communities~\cite{communities_notebook} dataset contains 128 columns. For both datasets, we duplicated the dataset multiple times (up to 10M rows for \airbnb and up to 100k rows for \communities) to investigate the effects of scaling  with the number of rows. After duplication, \airbnb exemplifies datasets with a moderate number of columns and a large number of rows, while \communities exemplifies those with a large number of columns. The upper limits on the two datasets cover around 98\% of the datasets in the UCI repository~\cite{uci}.

\stitle{Setup:} All of our experiments were conducted on a Macbook Pro with 32GB of RAM and an Intel Core i9 processor running macOS 10.15.6. The experiments were run using Python 3.7.7, pandas 1.2.1, and a version of lux-api 0.2.3 adapted for purpose of the experiments. We used \texttt{papermill}~\cite{papermill} to programmatically execute each notebook cell. We set $k$ for top k as 15 and apply \prune for any action where the number of visualizations exceeds $k$. For the sampling policy, we used cached random samples capped at 30k rows for approximating the visualization interestingness of dataframes over 30k rows\tr{ (the choice of this parameter is justified in Section~\ref{sec:exp_accuracy})}. For the runtimes reported, we exclude the frontend drawing time for each visualization given that it is constant and highly dependent on the chosen visualization library and frontend.
\cut{
    \stitle{Performance Metrics:}
    In \lux, the runtime is largely broken down into the time it takes to compute metadata ($t_{meta}$), to compute and generate visualization recommendations ($t_{rec}$), to render specified visualizations with a visualization library ($t_{render}$), and to display the widget on the frontend ($t_{frontend}$). For the purpose of the paper, we exclude $t_{frontend}$ given that the frontend drawing time for each visualization is highly dependent on the chosen visualization library and frontend.
}
\stitle{Conditions:} Our experiment measures the time it takes to execute every cell in the notebook across five different conditions: 
\change{
\begin{denselist}
    \item \textbf{no-opt}: Baseline condition with no optimization applied, representing a naive implementation of \lux where the results are explicitly computed at the end of every cell involving a reference to the dataframe\footnote{This condition is akin to the naive implementation in most visualization recommendation systems, where \tr{the }results are updated whenever the dataset is operated on.}. 
    \item \textbf{wflow}: Condition with the \wflow optimization applied.
    \item \textbf{wflow+prune}: Both \wflow and \prune applied. 
    \item \textbf{all-opt}: All \wflow, \prune, and \async applied, representing the best achievable performance.
    \item \textbf{pandas}: Condition with only \pandas and \textit{without} using \lux, representing the raw performance of dataframe workflows without the benefits of always-on visualizations. 
\end{denselist}
}
\tr{\subsection{Overall workflow performance (RQ1)}\label{sec:exp_overall}}
To evaluate the overall performance of \lux with a dataframe-based workflow, we measured the runtime for executing an example notebook involving \pandas.
\par \stitle{Workload:} The workload is based on publicly available notebooks on Kaggle for \airbnb and \communities. These notebooks follow a typical exploratory analysis of a dataframe that includes loading, transformation, cleaning, computing statistics, and machine learning. We modified these notebooks to print out dataframes and series at various points in the notebook akin to what a user would typically do for validating the results of operations. In addition, we label each cell in the notebook as either a print of a dataframe, print of a series, or neither (i.e., any non-\lux Python command) to separately measure the runtime for different cell types. Table~\ref{tbl:workload} shows the breakdown of the two notebook workloads by different cell types. We define \emph{overhead} as the difference in runtime between the all-opt and \pandas condition, i.e., the additional time required to support always-on visualizations via \lux. 
\begin{table}[h!]
    \vspace{3pt}
    \small
    \centering
    \caption{Table reports the number of cells for each type (N), the additional time incurred on top of \pandas for 10M \airbnb and 100k \communities (overhead), and the relative shape of the runtime distribution similar to Figure~\ref{fig:macrobenchmark},\ref{fig:print_df_perf}, (Distr.).} 
        \label{tbl:workload}
    \begin{tabular}{lccccc}
        & \multicolumn{2}{c}{\textbf{Airbnb}} & \multicolumn{2}{c}{\textbf{Communities}} &        \\ \cline{2-5}
        & N           & overhead [s]          & N              & overhead [s]            & Distr. \\ \hline
    Print df     & 14          & 21.18                 & 14             & 1.41                   & \includegraphics[height=1.5em]{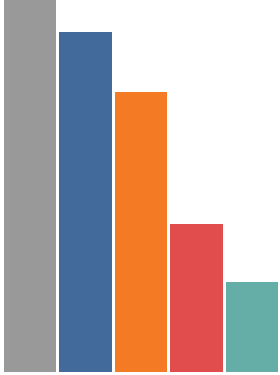}             \\ \hline
    Print Series & 7           & 0.61                  & 4              & 0.07                    & \includegraphics[height=1.5em]{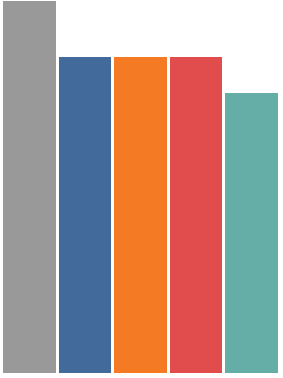}             \\ \hline
    Non-\lux  & 17          & 0                     & 25             & 0                       & \includegraphics[height=1.5em]{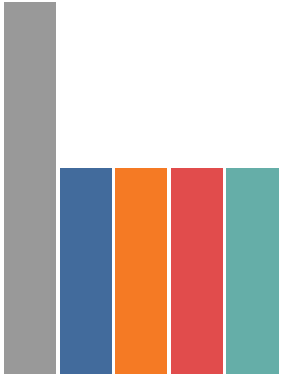}      
    \end{tabular}
\end{table}
\par \stitle{Overall runtime:} To understand the overall performance of \lux on dataframes with varying sizes, we varied the dataframe size from 10k to 10M rows. Figure~\ref{fig:macrobenchmark} displays the overall runtime  averaged over all cells in the notebook. We find that the best achievable performance with \lux led to significant speedup with up to 11X improvement in overall runtime for the \airbnb dataset (and up to 345X for \communities) compared to the no-optimization baseline. 
\begin{figure}[h!]
    \centering
    \includegraphics[width=0.9\linewidth]{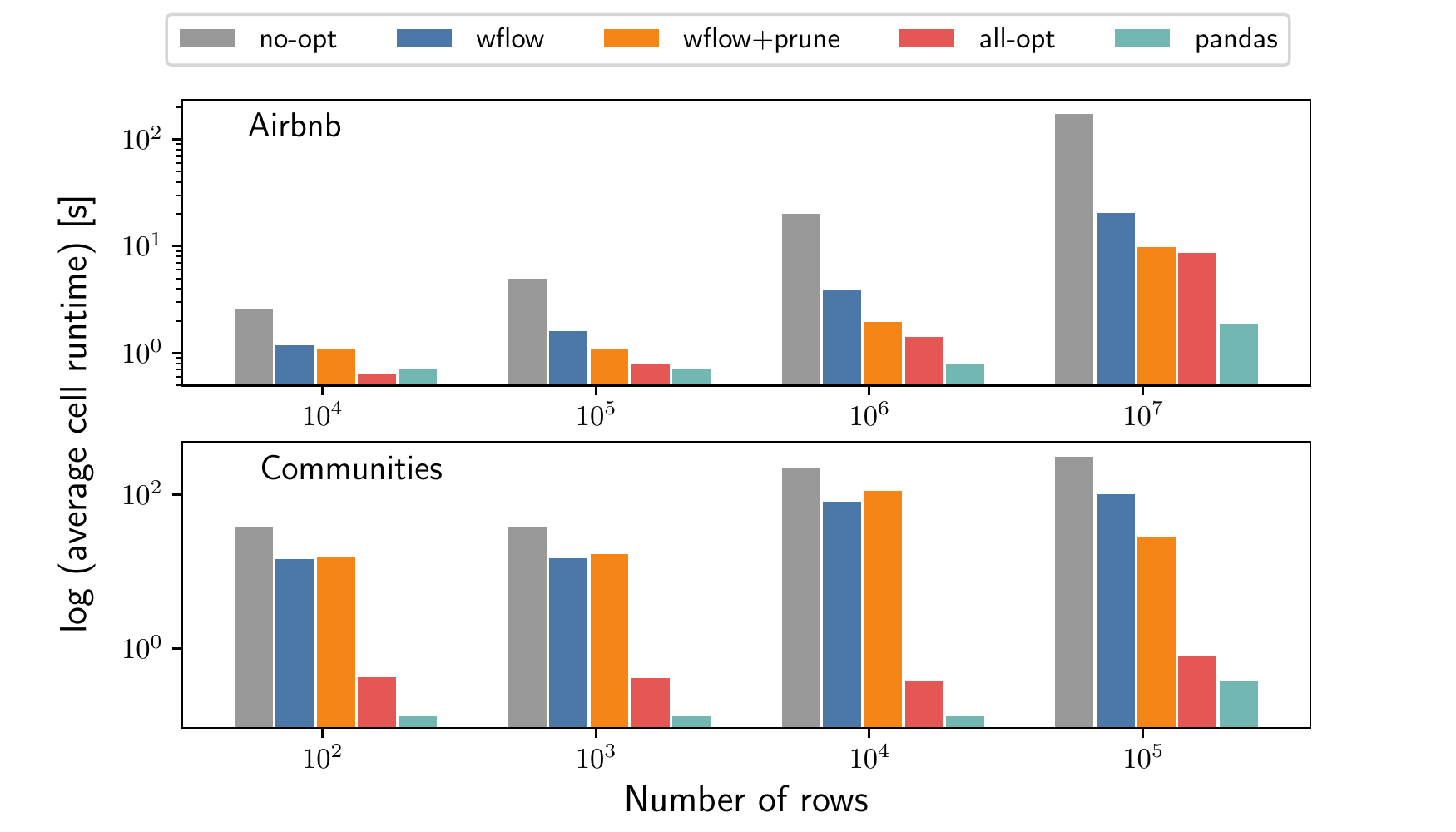}
    \caption{Average runtime of a notebook cell across the workload for different dataframe size and conditions.}
    \label{fig:macrobenchmark}
\end{figure}
\par \stitle{Printing dataframes and series:} We measure the performance of each cell that prints a dataframe or series to understand the overheads associated with \lux. Figure~\ref{fig:print_df_perf} shows the average time it takes for printing a dataframe for \airbnb and \communities. In particular, the overhead of \lux for each print can be determined by comparing against the cost for a print in \pandas. When the dataframe contains fewer than 1M rows for \airbnb, each print incurs no more than 2 seconds in addition to \pandas (in the 10M case, each print incurred an overhead of 21 seconds). For \communities, the overhead was no more than 1.5 seconds.
\par As shown in the sparkline visualization in Table~\ref{tbl:workload} row 2, the performance for printing series follows the same pattern as that of the dataframe. However, since series only involves a single column, it effectively avoids the costly procedure of traversing through a large search space. The overhead on top of \pandas is no more than 1 second for each series print even on the largest datasets.
\begin{figure}[h!]
    \centering
    \includegraphics[width=0.9\linewidth]{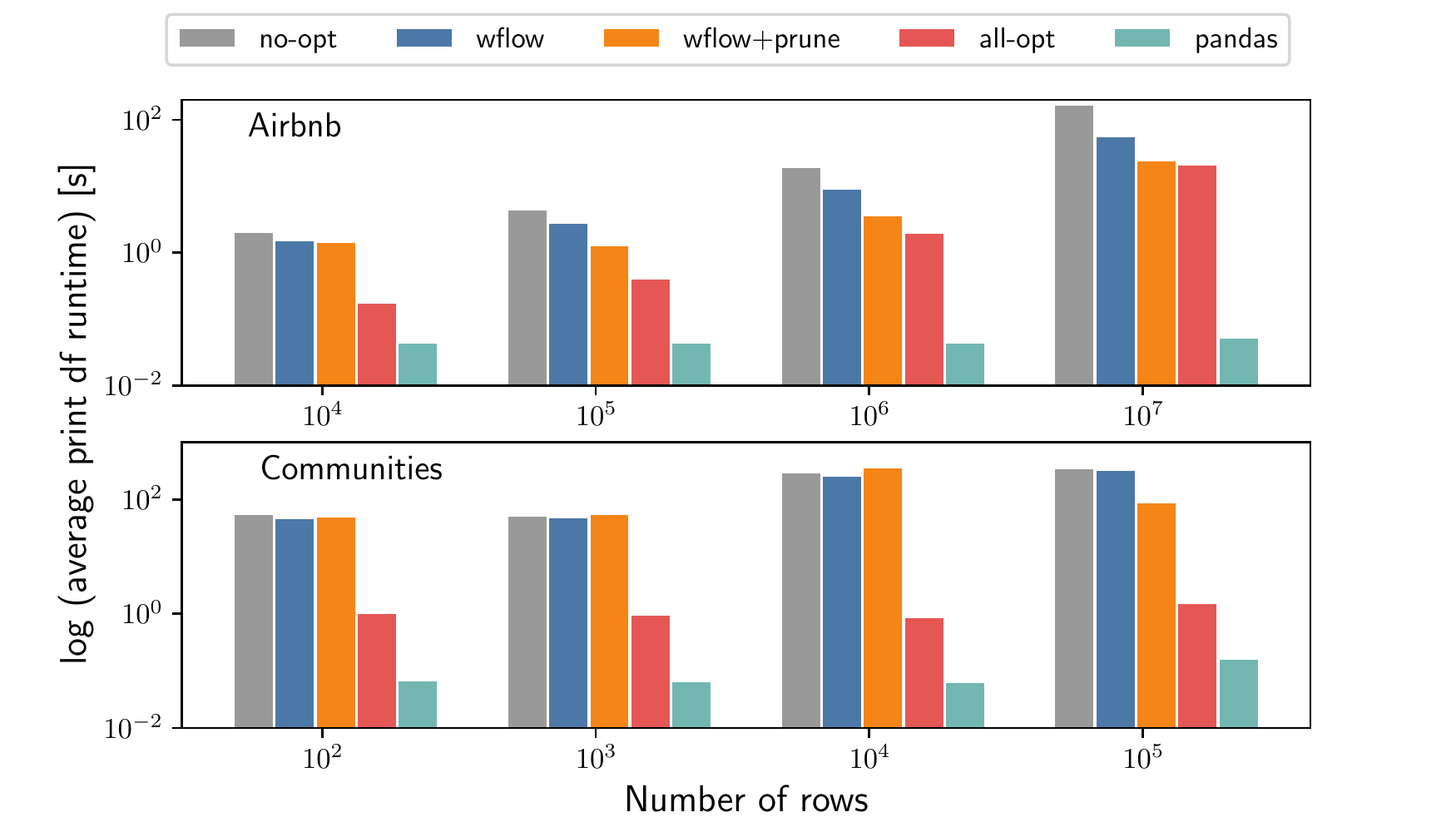}
    \caption{Average time for printing a single dataframe for different dataframe size and conditions.}
    \vspace{-5pt}
    \label{fig:print_df_perf}
\end{figure}

\par \stitle{Non-Lux operations:} Across all conditions except the baseline, the runtime for non-\lux operations (Table~\ref{tbl:workload} row 3) is the same---demonstrating how \lux incurs zero overhead on any Python operations in a notebook session. When compared against the baseline, \lux is over 100X faster for \change{10M} \airbnb and over 650X faster for \change{100k} \communities. The performance improvement for non-\lux operations demonstrates how \wflow's lazy evaluation strategy avoids unnecessary computation.
\tr{
    \subsection{Effect of dataframe width (RQ2)\label{sec:exp_ncols}}
    \par We investigate how the performance of \lux varies depending on the number of columns in the dataframe. 
    To understand the effect of the width of a dataframe ($w$), we measure the processing time for a single dataframe print (after the metadata has already been precomputed). Given the dependence of actions on data types, we leverage a synthetic dataset \tr{generated using the \texttt{faker}\cite{faker} library} to vary the number of columns in the dataframe, while fixing the proportion of data types. The simulated dataframe contains 100k rows with 78\% quantitative columns, 20\% nominal columns, and 2\% as temporal. Across the quantitative columns, half of the columns are integers, while the other half are floats. For the nominal columns, we generate columns of strings with varying cardinalities chosen based on a geometric series between 1 to 10000. 
    \par Figure~\ref{fig:ncols_accuracy_plots} left shows the runtime for different dataframe widths\footnote{We note that the no-opt condition is the same as \wflow in this case since we are only measuring a single print dataframe cell.}.  We note that the blue no-opt curve (power=2.53) scales exponentially with the number of columns. By applying the \prune and \async optimizations (red), \lux effectively lowers the cost of printing a dataframe by bringing the runtime closer to linear (power=1.07).
    \cut{
        We find that across all conditions the runtime for printing a dataframe scales exponentially with the number of columns. We perform least square fitting to determine the relationship between runtime and number of columns, which follows the functional form $t = a + b\cdot w^c$. Table~\ref{tbl:num_col_coeff} reports the fitting coefficients for different conditions. In particular, we note that by the all condtion changes the power coefficient ($c$) from a exponential to a close to linear relationship with respect to width. On the other hand, adding \prune also resulted in lower values of $b$, indicating a weaker dependence on the exponential term. The results indicate that \prune and \async enables better scaling for \lux on wide dataframes. \dtc{I don't think we need to understand the relationship between the runtime and the number of columns.}
        \begin{table}[]
            \begin{tabular}{llll}
                    & a     & b                    & c    \\ \hline
            \wflow (no-opt)                 & 14.22 & $2.28\times 10^{-4}$ & 2.53 \\
            \wflow + \prune                 & 8.15  & $5.52\times 10^{-8}$ & 3.76 \\
            \wflow + \prune + \async (all) & 0.37  & 0.03                 & 1.07
            \end{tabular}
            \caption{Table of fitting coefficient showing exponential dependence of runtime on dataframe width based on $t = a + b\cdot w^c$.}
            \label{tbl:num_col_coeff}
        \end{table}
    }
    \subsection{Effect on recommendation accuracy (RQ3)\label{sec:exp_accuracy}}
    To understand how the approximation-based \prune condition affects the recommended results, we experimented with different fractional sizes of the dataframe to be used in the sample and its effect on the recommendation ranking. 
    We compared the list of recommendations generated with and without the optimization applied. We computed Recall@15 of the top k results against the ground truth rankings. We chose recall, instead of other rank position-dependent measures, because the top-k visualizations are computed exactly and re-ranked after selection, so the metric only needs to capture how accurately the top-k visualizations are retrieved. 
    \cut{    
        \stitle{What actions is \prune applicable?} For \airbnb, \prune is only applicable for Correlation and Filter based on the criterion in Equation~\ref{eq:prune_condition}. Given that \communities has larger number of columns, the larger search space entails that \prune is more applicable to different actions.\dtc{lost in the last sentence} As shown in Figure~\ref{fig:ncols_accuracy_plots} center \dtc{what does each line mean in this figure}, across Correlation and Filter, we note that Filter requires more samples to achieve the same accuracy, since Filter enumerates over subsets of the data, it requires a larger overall sample to collect enough datapoints in the subsets.
        Overall, we find that actions where the variance in interestingness scores is low are more susceptible to errors, since data loss translate to ranking differences more readily. However, if the interestingness scores have a high spread, the rankings are more easily distinguishable, making room for more aggresive approximation strategy.
        \stitle{How big should the sample be?}
    }
    \par The recall curves in Figure~\ref{fig:ncols_accuracy_plots} right shows that for most actions 10\% (5k rows) is required in the sample for achieving over 90\% accuracy. For the 100k \airbnb dataset, the sample requirement is around 20-40\% (i.e., 20-40k rows). As a result, we chose the sampling cap in our experiment to be 30k rows to reach an average of 90.5\% on \airbnb dataset and near perfect ($\geq 95 \%$) on \communities. Compared to other actions, since Filter (light green in Figure~\ref{fig:ncols_accuracy_plots} right) enumerates over data subsets, it requires more samples to ensure enough data points per stratum to achieve the same accuracy.
    \cut{
        Correlation and Filter, we note that Filter requires more samples to achieve the same accuracy, since Filter enumerates over subsets of the data, it requires a larger overall sample to collect enough datapoints in the subsets.
        While Figure~\ref{fig:ncols_accuracy_plots} plots the accuracy curve against the sampling fraction, the factor that determines the sample accuracy is not only the fraction but the actual number of rows in the sample.\doris{are there AQP papers we can cite here or is this fairly obvious?} We experimented with the accuracy curve across different number of rows in the dataframe. To achieve a recall score of over 90\%, the \airbnb dataset required around 20-40k rows in the sample, while \communities required 5-7k rows. The difference in sample size required for attaining the accuracy is attributed to the interestingness characteristics of the dataset as outlined earlier. As a result, we chose the sampling cap in our experiment to be 30k rows to reach an average of 90.5\% on \airbnb dataset and near perfect ($\geq 95 \%$) on \communities.
    }
    \begin{figure}[h!]
        \centering
        \vspace{-2pt}
        \includegraphics[trim=0 1 0 5, clip=true, width=0.49\linewidth]{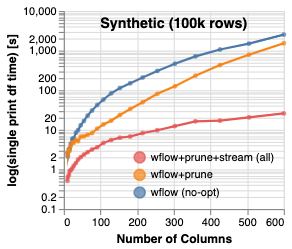}
        \includegraphics[trim=0 4 0 5, clip=true, width=0.49\linewidth]{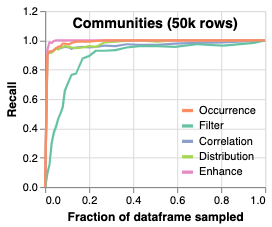}
        \caption{Left: Time spent for a single dataframe print varying the number of columns in a synthetic dataframe. 
                Right: Recall curve for different actions varying fractional samples of rows in the 50k \communities dataset.}
        \label{fig:ncols_accuracy_plots}
        \vspace{-8pt}
    \end{figure}
}


\change{
\section{Post-Deployment Evaluation}\label{sec:user}
\lux was first released in October 2020 and gained substantial traction in the open-source community since then.
In this section, we report on a field study with existing users of \lux and lessons learned from developing \lux.
\papertext{We summarize the key findings that would potentially be informative to future tool designers. Details of other post-deployment evaluations, including a first-use study, can be found in the technical report~\cite{lux-tr}.}
}
\tr{
    \subsection{First-use Controlled Study}

    We performed a  study to 
    understand participants' 
    initial impressions of \lux 
    and whether they are able to 
    use \lux effectively in 
    a controlled setting. 
    This study  was performed remotely from 
    October to November 2020 using lux-api 0.2.0. 
    This study was part of a 
    90-minute interactive session 
    where participants were first 
    introduced to the basics of \lux 
    and guided through a set of 
    hands-on exercises on how to use \lux. 
    The study was conducted with two focus groups:
    the first was a bootcamp for 
    industry data practitioners (N=20) 
    and the second was an online lecture 
    for students in a graduate-level 
    data visualization course (N=15). 
    Both groups engaged in the same 
    set of instructions and tasks. 
    The instructions and tasks 
    were made available to participants 
    via a web link to a live Jupyter notebook. 
    Participants were led through 
    three notebooks in sequence. 
    Each notebook contained examples 
    and exercises covering the key 
    concepts in \lux using three 
    datasets (College~\cite{college}, 
    Happy Planet Index~\cite{hpi}, and Olympics~\cite{medals}). Interactions on the \lux widget 
    and actions performed on the notebook 
    were logged via a custom extension~\cite{lux_logger}. 
    The session concluded with a 
    short survey documenting participants' experience. 
    Due to the remote and unsupervised study setting, not all participants submitted survey responses 
    or performed notebook operations that were logged. 

    \stitle{Study Findings.}
    We collected 16 survey responses
    (6 from bootcamp, 10 from lecture). 
    The results were thematically coded and 
    classified by one of the authors.
    In response to background questions 
    regarding the  
    existing exploration workflows 
    of the participants, 
    their concerns 
    echoed the pain points 
    that \lux aims to address, 
    including difficulty in 
    determining the ``right'' visualization 
    to plot (5/16), 
    modifying and iterating on visualizations (4/16), 
    and determining where to begin an analysis (4/16). 
    When asked to comment on aspects of \lux that they liked, 9/16 participants cited how the ability 
    to print and visualize dataframes was 
    the most useful. 
    Participants also noted how the 
    integration of \lux with their 
    data science workflow was seamless and intuitive. 
    When asked to comment on aspects of \lux 
    that they found challenging, 
    8/16 participants described 
    unfamiliarity and the learning curve 
    associated with the intent syntax. 
    When asked about what they 
    would like to see most in future versions, 
    participants were most interested in
    improving \lux's latency on large datasets 
    (12/16)\footnote{We note that the study was performed using the latest version of \lux at that point, which did not include many of the scalability improvements described earlier (\wflow was included, but not \async and \prune).}, 
    followed by support for 
    a wider and more useful 
    set of recommendations (8/16) 
    and making the intent language 
    more customizable (7/16). 
    At the end of the survey, 
    13/16 participants signed up for follow-ups 
    and expressed interest in continuing to use \lux.
    To evaluate whether participants 
    were able to accomplish controlled 
    tasks with \lux, 
    we collected 23 unique logs of 
    the participants' interaction 
    with the notebooks. 
    We qualitatively graded how well 
    participants performed across 
    the three exercises. 
    The task success rate for the 
    three exercises was 68\% (for composing 
    an intent indicating multiple views), 
    87\% (for specifying a desired \vis), 
    and 71\% (for creating a \vislist)\cut{\footnote{The scores discount users with no submitted responses.}}. 
    By inspecting the trace 
    of attempts, on average 
    participants were able to obtain 
    the first successful answer within 
    their first five tries. 
    Participants' most common mistakes 
    involved confusion around the syntax 
    for specifying multiple visualizations via union. 
    Finally, participants were 
    encouraged to try out one of the provided 
    datasets for open-ended exploration. 
    While participants successfully used 
    \lux to print and visualize their dataframes, 
    due to the setting and time constraints, 
    their interactions with \lux were brief. 
    The limited insight into how users 
    performs open-ended exploration with \lux 
    motivated the need for the following study. 
}
\tr{
    \begin{table*}
        \centering
        \scriptsize
        \resizebox{0.9\textwidth}{!}{%
        \begin{tabular}{llll}
            \toprule
            To what extent do you find the following functionalities in Lux useful?  &                 P1 &                 P2 &                 P3 \\
            \midrule
            Printing dataframe and inspecting recommended visualizations &        Very useful &        Very useful &   Extremely Useful \\
            Expressing analysis intent to steer recommendations &   Extremely Useful &   Extremely Useful &        Very useful \\
            Specifying visualization of interest via \vis &  Moderately useful &  N/A (Did not use) &        Very useful \\
            Specifying collections of visualizations of interest via \vislist &        Very useful &  N/A (Did not use) &        Very useful \\
            Exporting selected visualizations from Jupyter widget &        Very useful &   Extremely Useful &  N/A (Did not use) \\
            \midrule
            Lux makes it easier to ... &                 P1 &                 P2 &                 P3 \\
            \midrule
            Visualize my data across different stages in the data science workflow &              Agree &     Strongly agree &              Agree \\
            Plot a single visualization that I have in mind &     Strongly agree &     Strongly agree &     Strongly agree \\
            Identify what aspects of data I should visualize &     Strongly agree &              Agree &              Agree \\
            Determine what to do next in my exploration &              Agree &     Somewhat agree &            Neutral \\
            \bottomrule
        \end{tabular}
        }
        \vspace{5pt}
        \caption{Table of Likert scale ratings across the three field study participants.}
        \vspace{-15pt}
        \label{table:likertscale}
    \end{table*}  
}
\tr{
    \subsection{Field Study Interviews}
}
\papertext{
    \stitle{Field Study Interviews}
}
From December 2020 to January 2021,
we conducted semi-structured interviews 
with participants who used  
\lux in their data science work. 
We interviewed two industry 
data scientists in an insurance (P1) 
and retail company (P3), 
and a researcher in education (P2). 
Given that participants 
had extended exposure to \lux, 
our questions largely focused 
on understanding how \lux fits 
into their existing workflows. 
Before the interviews, 
participants used \lux over the 
span of 1-2 months 
in their professional data science work.\tr{ Their usage frequency varied: P1 used \lux daily, 
P2 used \lux once every one or two weeks, 
P3 used \lux around ten times in total. 
Unlike the first-use study 
where participants were led through 
instructions dedicated to how to 
create \vis and \vislist, 
field study participants learned 
how to use \lux on their own 
through tutorials and documentation
on our website.} We performed a walk-through of real-world
notebooks 
in which participants had used \lux. 

All three participants 
expressed that understanding 
their data was a challenge during exploration. 
In fact, two of the participants 
have developed their 
own homegrown solutions 
for past projects 
(echoing findings from Alspaugh et al.~\cite{alspaugh2019}), 
ranging from for loops 
across \mpl charts in notebooks 
to VBA scripts that generate plots in Excel. 
In their existing workflows, 
P1 and P2 visualized their 
data programmatically via \mpl, 
while P3 largely on 
Tableau's GUI for creating visualizations. 

\utitle{On dataframe visualizations:} 
All three participants 
expressed that they appreciated 
how the automatic visualizations 
provided by \lux afforded them 
quick insight into their dataframes 
without the need for code. 
P2 typically examines over 
100 columns of data as part 
of an educational course survey, 
and  stated that \lux sped up 
the amount of time for EDA 
by at least two-fold: \textit{`` it really helps speed up my exploratory analysis. If not, it will take me forever to go through these many variables.''} 
When asked about the scenarios 
for which they would toggle 
to the \lux view versus the default \pandas table, 
most participants preferred seeing 
the \lux view for the purposes of EDA. 
Participants described how 
they only use the \pandas table 
to quickly check 
if ``\textit{the data looks okay}'' (P1) 
and rarely toggle back to it 
unless they observe anomalous 
trends in the visualizations. 
During the study, P2 adopted 
a workflow 
where they sampled a single row 
to display the \pandas table in one notebook cell, 
then printed the \lux view 
in the cell below to check 
that the data falls in 
the expected ranges as displayed 
in the visualizations.

\utitle{On dataframe intents:} 
Participants indicated that 
the concept of intent 
was an intuitive way 
for steering the course of their analysis. 
P1 and P2 leveraged intent 
as a way of systematically 
exploring groups of variables 
they were interested in. 
To investigate their research questions, 
P2 listed groups of independent 
and dependent variables as their intent 
to explore each group one at a time. 
P1 and P3 used intent as a way 
of exploring predictive variables 
of interest, such as whether a customer 
purchased accessories alongside their orders, 
to help inform feature engineering 
for downstream machine learning. 
However, challenges in specification 
sometimes prevented them from making 
use of intent fully. 
In particular, P2 and P3 
both described that they were 
interested in exploring 
alternative data subsets 
for an attribute of interest 
(a query that is expressible 
in \lux's intent language); 
however, they were unaware 
that they could specify 
filter intent with wildcards. 
Improving the API for 
intent specification 
remains an important direction for future work. 

\tr{
    \utitle{On custom actions:} 
    Participants noted how the default \lux actions 
    largely covered the basic 
    sets of analyses that 
    they would typically perform on their own. 
    While most participants were unaware that \lux 
    supported the ability to create custom actions, 
    during various points in the interview, 
    they described additional actions 
    that they would find useful. 
    For example, P3 described how 
    they wanted to create a custom action 
    that lists the top ten dataframe columns with 
    the most influence over a desired predictive variable.
    Other participants described actions 
    that are similar to the default \lux actions, 
    but with a different ranking. 
    For example, P2 was interested 
    in categorical variables that involved 
    bar charts that looked very even, since that means that it has a closer-to-equal likelihood of being in either categories, so the trend is potentially interesting.
}

\utitle{On user-specified views:} 
\change{Somewhat surprisingly, the use of \vis and \vislist were} rarely brought up 
in the field study interviews. 
Possible explanations for their limited use 
include the unfamiliarity with these concepts 
and their usage of \lux in conjunction 
with other visualization tools. 
All participants used an existing 
visualization tool (e.g., matplolib or Tableau) 
while exploring their data with \lux. 
As a result, they simply used their 
familiar tools for specific visualizations 
when they knew exactly what to plot. 
To fully leverage \vis and \vislist in their work, 
participants often asked for ways to extend or customize the visualization type for a user-specified view.
For example, P3 explained how 
market share data was best visualized 
as a top-k pie chart, 
while P2 was interested in examining 
overlaid histogram distribution of 
different measures for binary variables, 
such as whether or not a course was open-ended. 
These findings indicate that increased flexibility 
in the intent language could afford the familiar visualization capabilities for users when creating specified views.

\utitle{Usage of \lux in data science workflows:} 
All three participants described 
using \lux explicitly 
in the exploration stage after 
data loading and cleaning, 
but before advanced analysis or modeling. 
P1 and P2 used \lux in conjunction with 
custom \mpl code that they repurposed 
for their analysis. 
When asked why participants 
did not print the dataframe 
for visualizations during the 
data transformation and cleaning phase, 
P1 and P3 answered that since 
the dataframe prints resulted 
in a few seconds of latency, 
they were hesitant to do it until 
they were ready to 
\textit{``chuck in [their] data and get the charts out''} (P3). Participants also described how \lux 
needed to be more robust in 
visualizing dirty or ill-formatted data.

\utitle{Summary and Limitations:} 
\tr{Table~\ref{table:likertscale} details 
participants' Likert scale ratings 
of the functionalities and benefits of \lux.} 
The average System Usability Scale (SUS)~\cite{sus} score across participants is 70/100. All three participants were interested in continuing to use \lux in their data science work. We learned that \vis and \vislist are 
not as discoverable and easy to use as 
always-on dataframe visualizations. Despite the enthusiasm around \lux,
we find that participants are still attached 
to their existing visualization tool for this functionality. 
They shared concerns around customizability 
and the inability to express 
their desired visualizations in \lux, 
pointing to the need for improving 
the flexibility of the intent language. 
Furthermore, a controlled study comparing \lux with a manual baseline approach would further quantify the expected benefits of the tool.

\tr{
    \subsection{Lessons from Developing \lux.}
}
\papertext{
    \stitle{Lessons from Developing \lux.}
}
We summarize the implementation challenges and lessons learned from the longitudinal open-source deployment of \lux over 15 months, with over 62k downloads.
Given that the nature of such engagement 
is largely organic, ranging from feature requests stemming from Github Issues to questions and discussions with users in our Slack community, these observations and findings will remain qualitative. 
\par \utitle{Metadata Propagation:} To preserve the comprehensive array of convenient operators 
offered by the dataframe API, we aimed to natively support any possible \pandas operations. This led to our design of \luxdf as a wrapper around the \pandas dataframe. However, dataframes can often get transformed to other intermediate data structures such as GroupBy, Series, or Index objects when users are working with dataframes. To this end, we extended specific \pandas functions and classes to ensure that the metadata and associated information is propagated across a workflow, so that the context does not get lost in intermediate operations. 
\par \utitle{Failproofing always-on dataframe display:} One of the reasons why dataframes have become popular is the ease and flexibility of being able to work with the data in a schema-free manner~\cite{modin-vision}. However, this can be problematic for generating recommendations, since underlying operations for visualization recommendations can lead to errors. For instance, dataframes can often be ill-formatted in a way that is not amenable to visualizations. One example of this is dataframes with missing values, or when dataframes contains mixed datatypes (a common issue when loading in spreadsheets).
To this end, we provide \pandas-consistent output behavior by safeguarding \lux with informative warnings, and falling back to \pandas upon internal errors, to always ensure that \lux provides at least the \pandas table as the default display. To allow users to effectively operate on their data, it is crucial that the system provides users with the \textit{unmodified}, \textit{consistent} state of the dataframe. In other words, \lux should not modify and change the user's dataframe in the process of visualization to maintain ``What You See Is What You Get'' (WYSIWYG) behavior. As a result, all recommendations in \lux are generated as views that are decoupled from the dataframe content.
\par \utitle{Integration with Downstream Reports:} One common use case for \lux is to get a quick overview of insights on a new dataset. We found that users often wanted a way to share their findings in their organization. Our initial use case for supporting downstream reporting was to allow users to select one or more visualizations and export it as \mpl or \code{altair} code. However, this approach quickly became unsustainable when users wanted to share many visualizations from their dashboard at the same time. To support presentation and collaboration, we implemented various options for export, from static HTML reports to integration with popular libraries for creating interactive ``data apps'', \papertext{such as}\tr{including DataPane~\cite{datapane}, Panel~\cite{panel}, and} Streamlit~\cite{streamlit}. Given that many \lux users often share their findings with business stakeholders without a Python development setup, future work might include supporting exports to readily-accessible presentation formats, such as PDF or Powerpoint.
\par Another lesson that we learned is that ease of initial installation and setup is a primary driver impacting the adoption of tools like \lux.  In a similar vein, our user surveys and online discussions suggest that the minimal API\tr{ as demonstrated in our documentation and tutorials} is attractive to many data practitioners.
\section{Conclusion}\label{sec:conclusion}
\change{We propose \lux, an always-on visualization framework for exploratory dataframe workflows. \lux is a lightweight wrapper that reduces the barrier of visualizing data, enabling seamless exploration and visual discovery in-situ. 
To provide better visualization recommendations, we make use
of user-provided} intent and history, as well as structural information and metadata. 
\change{We extend and evaluate various optimization strategies that minimize the overhead of \lux, including approximate query processing, lazy computation, and caching and reuse}. \lux's adoption over the last year and success of user evaluation points to its importance for \change{dataframe workflows}---steering users towards valuable insights as they ponder what to do next 
with their data.

\begin{acks}
\noindent We thank the anonymous reviewers for their valuable feedback. 
This work is supported by a Facebook Fellowship, grants IIS-2129008, IIS-1940759, and IIS-1940757 awarded by the National Science Foundation, and funds from the Alfred P. Sloan Foundation. The content is solely the responsibility of the authors and does not necessarily represent the official views of the funding agencies and organizations.
\end{acks}

\bibliographystyle{abbrv}
\bibliography{refs}

\end{document}